\documentclass[12pt]{article}
\usepackage{lipsum}
\usepackage{setspace} 
\usepackage{amsmath, amsfonts, amssymb, amsthm, newlfont, setspace, graphicx, rotating, newtxmath, mathrsfs}
\usepackage{lscape}
\usepackage{multicol}
\usepackage{xcolor}
\usepackage{booktabs, siunitx}
\usepackage[normalem]{ulem}
\usepackage[colorlinks=true, linkcolor=blue, citecolor=blue, urlcolor=blue]{hyperref}
\usepackage{bbm}
\usepackage[english]{babel}
\usepackage[autostyle]{csquotes}
\usepackage{dsfont}
\usepackage[svgnames,table]{xcolor}
\usepackage[tableposition=above]{caption}
\usepackage{multirow}
\usepackage[style=apa, backend=biber, citestyle=authoryear, sorting=nyt]{biblatex}
\DeclareLanguageMapping{american}{american-apa}
\usepackage{csquotes}

\newtheorem{theorem}{Theorem}
\newtheorem*{theorem*}{Theorem}
\numberwithin{equation}{section}
\newtheorem{corollary}{Corollary}[theorem]
\newtheorem{remark}{Remark}

\usepackage{mathtools}  
\usepackage{diffcoeff}
\usepackage{algorithm}
\usepackage{algpseudocode}
\usepackage{enumitem}
\usepackage[margin=1in]{geometry}
\usepackage{fancyvrb}
\usepackage{tikz} 
\usetikzlibrary{shapes.geometric, arrows, arrows.meta, shadows, shadows.blur, positioning, calc, decorations.pathmorphing}
\usepackage{stackengine}
\usepackage{float}
\usepackage{placeins}
\usepackage{afterpage}
\usepackage{graphicx}
\usepackage{wrapfig}
\usepackage{ragged2e}
\usepackage{mdframed}
\usepackage{accents}
\usepackage{chngcntr} 
\counterwithin{figure}{section} 
\usepackage[x11names]{xcolor}
\usepackage{appendix} 
\begin{document}
\title{A Robust Persistent Homology : Trimming Approach}
\author{\small 
	Tuhin Subhra Mahato \\
	\small IIT Kanpur\\
	\small Department of Mathematics and Statistics \\
	\small  Kanpur 208016, India\\
	{\small Email: tuhinsm23@iitk.ac.in }\\
	\and
	\small Subhra Sankar Dhar \\
	\small  IIT Kanpur\\
	\small   Department of Mathematics and Statistics \\
	\small Kanpur 208016, India\\
	{\small Email: subhra@iitk.ac.in}\\
}
\date{}
\maketitle
\begin{center}
    \textbf{Abstract} 
\end{center} 
This article studies the robust version of persistent homology based on trimming methodology to capture the geometric feature through support of the data in presence of outliers. Precisely speaking, the proposed methodology works when the outliers lie outside the main data cloud as well as inside the data cloud. In the course of theoretical study, it is established that the Bottleneck distance between the proposed robust version of persistent homology and its population analogue can be made arbitrary small with a certain rate for a sufficiently large sample size. The practicability of the methodology is shown for various simulated data and bench mark real data associated with cellular biology.


\section{Introduction}
Topological Data Analysis (TDA) has emerged as a powerful framework in modern statistics and data science for uncovering the shape of data, that is, the geometric and topological structure of the support of an unknown probability distribution based on finite samples. The ability to extract structural information directly from data makes TDA well-suited for analyzing data that arises across a diverse range of domains, including biological sciences, \textcite{nicolau2011topology}; materials science, \textcite{hiraoka2016hierarchical}; astronomy, \textcite{pranav2017topology}; signal processing, \textcite{robinson2014topological}; financial time series analysis, \textcite{gidea2018topological}; climate modelling, \textcite{reininghaus2015stable}; sensor networks, \textcite{de2007homological}, and many others. In the context of effect of outliers in the data structure, like parametric model, a single outlier in the sample can also distort the support of the underlying distribution by creating artificial topological features or filling in genuine cavities. Thus, contamination not only introduces noise but also alters the topology of the shape of the data, thereby making it difficult to infer. This motivates us to develop robust methods of topological inference.
\subsection{Literature Review and Contribution}
The problem of support estimation has received attention in statistical learning and nonparametric inference. Estimating the support of an unknown distribution is a fundamental problem in statistics, with applications in anomaly detection, clustering, and manifold learning. One of the earliest and simplest estimators is the empirical support, defined as the set of observed sample points $\mathbb{X}_n=\{X_1,\cdots, X_n\}$. While intuitive, this approach does not capture geometric or topological features of the data. To improve upon this, \textcite{devroye1980detection} proposed an estimator constructed by the union of closed balls, and 
under some regularity conditions, this estimator is consistent for the true support.
However, classical methods often fall short when the data lie on lower-dimensional structures embedded in high-dimensional spaces. In those situations, such methods are inadequate in capturing topological features like loops and voids. This limitation leads to the domain of TDA (\textcite{carlsson2009topology}), and it provides a way to infer topological features of the support of an unknown distribution using tools from algebraic topology, such as homology and persistence homology (see \textcite{balakrishnan2012minimax} and references therein).

Parallel to these developments on homology estimation, attention shifted towards persistence diagrams, which encapsulate the multiscale topological features through persistent homology. The statistical theory of persistence diagrams was advanced through the work of \textcite{mileyko2011probability}, who studied probability measures on the space of persistence diagrams, and \textcite{fasy2014confidence}, who introduced confidence sets for persistence diagrams.
A major contribution to the statistical consistency of persistence-based summaries came from \textcite{chazal2015convergence}, who established convergence rates for persistence diagram estimation.
To make persistence-based summaries easier for statistical analysis, researchers developed stable, vectorized representations of persistence diagrams, including Persistence Landscapes (\textcite{bubenik2015statistical}) and Persistence Images (\textcite{adams2017persistence}). Recently, 
 \textcite{kumar2022testing} introduced testing procedures based on Betti numbers to compare homological structures across different datasets. In a follow-up study, through the lens of persistent homology, they extended these ideas to functional data analysis by identifying significant structural patterns in smooth regression curves (see, for instance, \textcite{kumar2023novel}). By preserving essential geometric and topological information, these advances connect persistent homology and diagram with statistics, bringing them closer to real-data analysis.

Despite substantial progress in the development of theoretical and algorithmic tools in TDA, many of its statistical aspects remain relatively underexplored. A key challenge in the statistical TDA is estimating topological features of the support of an unknown probability distribution based on a finite dataset, particularly in the presence of contamination. To address this issue, several robust methods have been proposed in recent literature; see, for example, \textcite{bobrowski2017topological}, \textcite{chazal2018robust}, \textcite{dakurah2025maxtda}. One such effective method can be data trimming, which discards a small fraction of observations with unusually large or small average pairwise distances to the rest of the observations, often regarded as outliers. Here the form of the distance depends on the distance metric related to space containing data. Now, in the presence of outliers or influential and/or extreme observations, after removing these outliers, the trimmed data set is expected to recover the actual topological features inherent in the data. 
 Consequently, persistent homology computed over this trimmed support (i.e., support of the trimmed observations) yields a persistence diagram that reveals ``true" topological structure of the data. In this way, the proposed method based on persistent homology enables reliable identification of stable and interpretable persistent topological features for contaminated data.

\subsection{Challenges}
Developing a robust trimming-based framework for TDA raises several challenges. First, the theoretical analysis must account for the geometric complexity of the underlying support, which may include non-manifold and other irregular structures. At the same time, it must handle the statistical dependence induced by data-driven trimming. Second, the choice of trimming proportions leads to a bias–variance trade-off. Insufficient trimming can preserve contamination-induced bias, whereas excessive trimming can induce loss of genuine signal. Our work addresses these challenges by (i) establishing uniform convergence bounds under some regularity conditions, and (ii) providing guidance for the selection of trimming parameters, validated through simulation studies and real data applications.

A key ingredient in our analysis, used to address the dependence induced by data-driven trimming, is a probabilistic coupling principle introduced later in Assumption (A3). This assumption ensures the existence of an independent surrogate sample that approximates the trimmed dataset up to a small coupling discrepancy. Analyzing this independent surrogate sample allows us to apply standard concentration and covering arguments. The resulting bounds are then transferred back to the original, dependent trimmed sample with an additional error that is explicitly controlled by the coupling discrepancy.

\subsection{Organization} 
 The rest of the article is organized as follows. \hyperref[section 2]{\textcolor{blue}{Section \ref{section 2}}} outlines the fundamental concepts in statistical TDA that motivate our study, focusing on persistent homology and the persistence diagram. \hyperref[section 3]{\textcolor{blue}{Section \ref{section 3}}} provides a detailed explanation of the proposed methodology. The results and main findings are discussed in \hyperref[section 4]{\textcolor{blue}{Section \ref{section 4}}}. \hyperref[section 5]{\textcolor{blue}{Section \ref{section 5}}} addresses the data driven choice of trimming proportions. \hyperref[section 6]{\textcolor{blue}{Section \ref{section 6}}} evaluates the empirical behaviour of the proposed method through simulation studies, while \hyperref[section 7]{\textcolor{blue}{Section \ref{section 7}}} illustrates the practical relevance of our approach by applying it to a real-world dataset. Finally, in \hyperref[section 8]{\textcolor{blue}{Section \ref{section 8}}}, we provide some concluding remarks, and the proof of Theorem 1 is provided in the \hyperref[Appendix]{\textcolor{blue}{Appendix}}.
\section{Background}\label{section 2}
\subsection{Metric and Probabilistic Framework}\label{section 2.1}
Let $\mathcal{M}$ be a compact metric space equipped with a distance function $d:\mathcal{M}\times\mathcal{M}\to [0,\infty)$.
The class $\mathscr{P}_{a,b}(\mathcal{M})$ consists of all probability measures $\mathbb{P}$ on $\mathcal{M}$ satisfying the \textbf{$(a,b)$-standard assumption}: there exist constants $a, b > 0$, and $\varepsilon_0 \in (0,1)$, such that
\begin{equation}\label{eq:2.1}
    \mathbb{P}(B_\varepsilon(x)) \geq \min(1, a \varepsilon^b),~~ \text{for all}~ x \in \textbf{Supp}(\mathbb{P}),~ 0 < \varepsilon < \varepsilon_0,
\end{equation}
where $B_\varepsilon(x):= \{y \in \mathcal{M}: d(x, y)< \varepsilon \}$  denotes the open ball of radius $ \varepsilon $ centered at $x$, and \textbf{Supp($\mathbb{P}$)} = $\{x \in \mathcal{M}: \mathbb{P}(B_\varepsilon(x))>0, ~ \text{for all}~~ \varepsilon>0\}$ denotes the support of $\mathbb{P}$.
Let $\mathbb{P} \in \mathscr{P}_{a,b}(\mathcal{M})$  be a probability measure with compact support \textbf{Supp($\mathbb{P}$)} $\subset $~$\mathcal{M}$. 
The support \textbf{Supp($\mathbb{P}$)} is assumed to satisfy a uniform interior ball condition at some scale $r>0$, that is, for all $x \in \textbf{Supp}(\mathbb{P})$, there exists $y \in \textbf{Supp}(\mathbb{P})$ such that $x \in \bar{B}_r(y)\subset \textbf{Supp}(\mathbb{P})$. In addition, the boundary of \textbf{Supp($\mathbb{P}$)} is assumed to be of class $C^2$ (twice continuously differentiable), ensuring bounded curvature and excluding regions with arbitrarily small radius of curvature.

Moreover, the lower mass-bound condition (see \eqref{eq:2.1}) ensures that $\mathbb{P}$ does not have negligible probability mass in arbitrarily small neighbourhoods around points in its support. This regularity condition plays a crucial role in establishing uniform convergence in support estimation and stability of topological descriptors. These assumptions are widely adopted in the literature of nonparametric support estimation. See, for instance, the works of  \textcite{tsybakov1997nonparametric}, \textcite{cuevas2007nonparametric}, \textcite{genovese2012minimax}, \textcite{chazal2011geometric}, \textcite{aamari2019estimating}.
\subsection{Topological Spaces and Homeomorphism}
In TDA, the focus shifts from pairwise distances between data points to the shape or connectivity of data. Therefore, it is important to study the shape of $\textbf{Supp}(\mathbb{P})$ through the lens of topology. To capture the abstract notion of shape, we consider a topological space and a few associated issues.
\subsubsection{Topological Space}
A topological space is a pair $(\mathcal{X},\tau)$, where $\mathcal{X}$ is a set and $\tau$ is a collection of subsets of $\mathcal{X}$ satisfying the following axioms:
\begin{enumerate}
    \item $\emptyset, \mathcal{X} \in \tau$,
    \item If $\{U_i\}_{i \in I} \subseteq \tau$, then $\bigcup_{i\in I} U_i \in \tau$,
    \item If $U_i\in \tau$ for $i=1, \cdots,n$, then $\bigcap_{i=1}^{n}U_i\in \tau$.
\end{enumerate}
In our framework, the ambient space $\mathcal{M}$ is endowed with a metric $d$, and hence, it carries a metric topology $\tau_d$ generated by open balls $B_\varepsilon(x):=\{y\in \mathcal{M}: d(x,y)<\varepsilon\}$. This makes $(\mathcal{M}, \tau_d)$ a topological space with the geometric structure introduced in \hyperref[section 2.1]{\textcolor{blue}{Section \ref{section 2.1}}}.
\subsubsection{Homeomorphism}
 A function $f: (\mathcal{X},\tau_\mathcal{X}) \to (\mathcal{Y},\tau_\mathcal{Y})$ between two topological spaces $\mathcal{X}$ and $\mathcal{Y}$ is called a homeomorphism if it satisfies the following conditions:
\begin{enumerate}
    \item $f$ is one-to-one and onto.
    \item $f$ is continuous: for every open set $V \in \tau_\mathcal{Y}$, the pre-image $f^{-1}(V) \in \tau_\mathcal{X}$.
    \item $f^{-1}: \mathcal{Y} \to \mathcal{X}$ is continuous.
\end{enumerate}
If such a function exists, then $\mathcal{X}$ and $\mathcal{Y}$ are said to be homeomorphic to each other. Homeomorphic spaces share the same connectivity properties, number and types of holes, and other topological invariants. 

From a statistical perspective, the goal is to infer the topological features of $\textbf{Supp}(\mathbb{P})$ from finite samples. Homeomorphism ensures that the inferred topological features are invariant under continuous transformations, such as stretching, bending, twisting, without tearing or glueing. For instance, whether the sample lies on a circle or any closed curve embedded in higher dimensions, the existence of a homeomorphism guarantees that the fundamental cycle is preserved. This motivates the use of simplicial homology, which works directly on simplicial complexes and serves as the backbone of topological data analysis. 
\subsection{Simplicial Homology}
Consider an observed dataset $\mathbb{X}_n=\{X_1, X_2,\cdots, X_n\}\subset \mathbb{R}^{m}$, which is assumed to be an i.i.d. sample from an unknown distribution $\mathbb{P}$. The goal is to estimate the topological features of the support $\textbf{Supp}(\mathbb{P})$ of the underlying distribution $\mathbb{P}$ using the support $\textbf{Supp}(\mathbb{P}_n)$ of the empirical measure $\mathbb{P}_n$, where $\mathbb{P}_{n}$ is a suitable estimator of $\mathbb{P}$. Since $\textbf{Supp}(\mathbb{P})$ is not directly observable, we approximate its topological structure using  simplexes based on $\mathbb{X}_n$. Next, the details on simplex is provided. 
\subsubsection{Simplex} 
Let $\mathbb{X}_n=\{X_1, X_2,\cdots, X_n\}\subset \mathbb{R}^{m}$ be a finite set of ponts. A subset $\{X_{i_0},X_{i_1},\cdots ,X_{i_k}\} \subseteq \mathbb{X}_n$ forms a $k$-simplex if the points are affinely independent, that is, the vectors $\{X_{i_1}-X_{i_0}, X_{i_2}-X_{i_0}, \cdots, X_{i_{k}}-X_{i_0}\}$ are linearly independent in $\mathbb{R}^{m}$. The $k$-simplex $\sigma_k=[X_{i_0},X_{i_1},\cdots,X_{i_k}]$ spanned by these points is the set of all convex combinations of these points:
\begin{equation*}
    \sigma_{k}=\Bigg\{ \sum \limits_{j=0}^{k} \alpha_{j} X_{i_j} \mid X_{i_j}\in \mathbb{X}_n,~\alpha_{j} \geq 0\quad\text{and}\quad \sum \limits_{j=0}^{k} \alpha_{j}=1\Bigg\}.
\end{equation*}
A  $0$-simplex corresponds to a point, a $1$-simplex to an edge, a $2$-simplex to a filled triangle, and more generally, a $ k$-simplex is the convex hull of $(k+1)$ affinely independent points. 
\subsubsection{Simplicial Complex}
A simplicial complex $\mathcal{K}$ is a finite collection of simplices $\sigma_i$, $i\in I$ satisfying the following conditions:
\begin{enumerate}[label=(\arabic*)]
    \item  If $\sigma_i \in \mathcal{K},$  and  $\sigma_j \subset \sigma_i$ then $ \sigma_j \in \mathcal{K}$; $\forall ~i,j\in I$.
    \item For  $\sigma_i, ~\sigma_j \in \mathcal{K}$ with $i \neq j $, either $\sigma_i\cap \sigma_j = \emptyset $ or, $\sigma_i\cap \sigma_j = \sigma_k \in \mathcal{K} $, where $\sigma_k\subset \sigma_i,~\sigma_k\subset\sigma_j$ and $i,j,k\in I$.
\end{enumerate}

 An abstract simplicial complex $\Delta$ over $\mathbb{X}_n$ is a collection of finite, non-empty subsets of $\mathbb{X}_n$ such that:
\begin{enumerate}[label=(\arabic*)]
    \item $\{X_i\} \in \Delta$, for all $X_i \in \mathbb{X}_n$.
    \item If $\sigma \in  \Delta$ and $\tau \subset \sigma$ then $\tau \in  \Delta$.
\end{enumerate}
This combinatorial description is essential in data analysis, where geometry may be unknown but pairwise distances are available.
\subsubsection{Construction of Simplicial Complex from Data}
A natural way to build a simplicial complex $\mathcal{K}$ from a finite sample is by connecting points based on their pairwise distance induced by distance function $d(., .)$, and suppose that the scale parameter $\varepsilon >0$ determines how close points need to be to form connections. Among the simplicial complexes available in the literature, Vietoris-Rips Complex (denoted by $\mathbb{VR}_\varepsilon\big(\textbf{Supp}(\mathbb{P}_n)\big)$) is one of the wide used simplicial complex based on the empirical measure $\mathbb{P}_{n}$ and the scale parameter $\varepsilon$.   

\textbf{Vietoris-Rips Complex:}
Given a finite sample $\mathbb{X}_n=\{X_1, X_2,\cdots, X_n\}\subset \mathbb{R}^{m}$, the Vietoris Rips complex $\mathbb{VR}_\varepsilon\big(\textbf{Supp}(\mathbb{P}_n)\big)$ constructed over $\textbf{Supp}(\mathbb{P}_n)$ at scale $\varepsilon >0$ is defined as follows: for any integer $k\geq 0$, a collection of $(k+1)$ points \(\{X_{i_0}, X_{i_1}, \dots, X_{i_k}\} \subset \textbf{Supp}(\mathbb{P}_n)\) forms a $k$-simplex $\sigma_{k}=[X_{i_{0}},\cdots,X_{i_{k}}]$ in $\mathbb{VR}_\varepsilon\big(\textbf{Supp}(\mathbb{P}_n)\big)$  if every pair of points in the set satisfies $d(X_{i_p},X_{i_q})\leq 2\varepsilon, ~ \forall~ 0 \leq p,q \leq k$.
\subsubsection{Geometric Realization}
In many real-life scenarios, $\mathbb{X}_n$ is drawn from metric spaces, which are not naturally embedded in Euclidean space, such as manifolds (e.g., a smooth curved surface). For such cases, to visualize the simplicial complex and to compute topological invariants such as Betti numbers, it is often necessary to map the simplicial complex  $\mathcal{K}$ into Euclidean space. This process is known as the geometric realization of a simplicial complex, and the resulting embedded simplicial complex is called a geometric simplicial complex, denoted as  $|\mathcal{K}|$. Technically speaking, a geometric realization of a $k$-simplex $\sigma_{k}=[X_{i_{0}},\cdots,X_{i_{k}}]$ is defined as follows:
\begin{equation*}
    |\sigma_{k}|= \Bigg\{ \sum \limits_{j=0}^{k} \alpha_{j}y_{j} \in \mathbb{R}^m \mid  y_{j}\in \mathbb{R}^m, ~ \alpha_{j} \geq 0\quad \text{and}\quad \sum \limits_{j=0}^{k} \alpha_{j}=1\Bigg\}.
\end{equation*}
Thus, geometric realization serves as a bridge between abstract topology and data analysis, especially when data are embedded in non-Euclidean spaces.
\subsubsection{Triangulation}
Geometric realization provides a way to embed abstract simplicial complexes into Euclidean space. However, when data are sampled from an unknown topological space, beyond visualization, the aim extends to accurately recover the topological features of the space. This leads to the notion of triangulation. A triangulation of a topological space $\mathcal{X}$ is a simplicial complex whose geometric realization is homeomorphic to $\mathcal{X}$. In notation, triangulation of a topological space $\mathcal{X}$ is a pair ($|\mathcal{K}|,~ h$), where $|\mathcal{K}|$ denotes a simplicial complex in $\mathbb{R}^m$ and $h:|\mathcal{K}| \rightarrow \mathcal{X}$ is a homeomorphism. In this way, triangulation allows us to represent a topological space by a simplicial complex. Not every topological space is triangulable. Although the spaces encountered in data analysis typically belong to classes that admit triangulations. The next step is to extract information about topological features from the combinatorial structure of the simplicial complex. This is achieved through homology theory, which assigns to each simplicial complex a sequence of algebraic objects (called homology groups) that capture different-dimensional topological features.
\subsubsection{Homology}
Let $|\mathcal{K}|$ be a geometric simplicial complex. The simplicial homology group $H_k(|\mathcal{K}|)$ provides the $k$-dimensional topological features of $|\mathcal{K}|$ in terms of abelian groups. The construction of homology proceeds as follows:
\paragraph{\textbf{Oriented $k$-simplex:}} Consider a $k$-simplex $\sigma_{k}=[X_{i_{0}},\cdots,X_{i_{k}}]$. An ordered $k$-simplex is a tuple $(X_{\pi_{(0)}},\cdots,X_{\pi_{(k)}})$, where $\pi \in S_{k+1}$ is a permutation of the indices $\{{i_{0}},\cdots,{i_{k}}\}$, and the set $S_{k+1}$ represents all possible orderings of the vertices of $\sigma_{k}$. Two orderings $\big(X_{i_{(0)}},\cdots,X_{i_{(k)}}\big)$ and $\big(X_{\pi_{(0)}},\cdots,X_{\pi_{(k)}}\big)$ of the vertices of $\sigma_{k}$ are said to have the same orientation if $\pi \in S_{k+1}$ is an even permutation. An oriented $k$-simplex is then the equivalence class of orderings of $(k+1)$ vertices under even permutation.
\begin{equation*}
   [X_{i_{(0)}},\cdots,X_{i_{(k)}}] :=\bigg\{[X_{\pi_{(0)}},\cdots,X_{\pi_{(k)}}]:  \pi \in S_{k+1}~~ \text{is even}\bigg\}.
\end{equation*}
 Geometrically, orientation insights towards the notion of direction. For instance, $[X_{i_{(0)}},X_{i_{(1)}}]$ denotes a directed edge from $X_{i_{(0)}}$ to $X_{i_{(1)}}$, while $[X_{i_{(1)}},X_{i_{(0)}}]$ = - $[X_{i_{(0)}},X_{i_{(1)}}]$ represents the opposite direction.
\paragraph{\textbf{$k$-chain and Chain Group:}} Once the orientation of simplices is fixed, one can form chains by combining them algebraically.
A $k$-chain is a linear combination of oriented $k$-simplices in $|\mathcal{K}|$ with coefficients in $\mathbb{Z}_2=\{0,1\}$. For each non-negative integer $k$, the set of all such $k$-chains forms an abelian group under addition modulo 2, denoted by $\mathbb{C}_{k}(|\mathcal{K}|)$. This group is called the $k$-th chain group, and is generated by the oriented $k$-simplices in $|\mathcal{K}|$.
\begin{align*}
     \mathbb{C}_{k}(|\mathcal{K}|)= \Bigg\{\sum \limits_{i=1}^{N} c_{i}~ \sigma_{k}^{(i)}\mid c_{i} \in \{0,1\},~ \sigma_k^{(i)}~ \text{is an oriented $k$-simplex in}~ |\mathcal{K}|\Bigg\},
\end{align*}
where $N$ is the number of $k$-simplices in $|\mathcal{K}|$, and $\sigma_k^{(i)}$ denotes the $i$-th oriented $k$-simplex.
\paragraph{\textbf{Chain Complex:}}
The next step is to connect the chain group across dimensions. This is done through the boundary operator $\partial_{k}: \mathbb{C}_{k}(|\mathcal{K}|)\rightarrow \mathbb{C}_{k-1}(|\mathcal{K}|)$, defined on an oriented $k$-simplex $[X_{i_{0}},\cdots,X_{i_{k}}]$ as
\begin{equation*}
    \partial_{k}\big([X_{i_{0}},\cdots,X_{i_{k}}]\big) :=\sum_{j=0}^{k}(-1)^j [X_{i_{0}},\cdots,\hat{X}_{i_{j}},\cdots,X_{i_{k}}],
\end{equation*}
where $\hat{X}_{i_{j}}$ denotes the omission of the vertex $X_{i_{j}}$. Thus boundary operator maps on $k$-chains in a simplicial complex $|\mathcal{K}|$ to $(k-1)$-chains in $|\mathcal{K}|$. The collection of chain groups together with the boundary maps $\partial_{k}$, forms a chain complex $\Big\{ \mathbb{C}_{k}(|\mathcal{K}|),\partial_k \Big\}_{k \geq 0}$, which is as follows. 
\begin{center}
     $\cdots \xrightarrow{\partial_{k+1}} \mathbb{C}_{k}(|\mathcal{K}|) \xrightarrow{\partial_{k}} \mathbb{C}_{k-1}(|\mathcal{K}|) \xrightarrow{\partial_{k-1}} \cdots \xrightarrow{\partial_{2}}\mathbb{C}_{1}(|\mathcal{K}|) \xrightarrow{\partial_{1}} \mathbb{C}_{0}(|\mathcal{K}|) \rightarrow 0,$
\end{center}
with the property $\partial_{k}\circ \partial_{k-1}=0$ for all $k \geq 1$. This ensures that the boundary of a boundary is always empty, which is a cornerstone idea in homology.
\paragraph{\textbf{Cycles and Boundaries:}} With the chain complex in hand, one can now able to distinguish between two fundamental subspaces of chains: cycles and boundaries.

The space of $k$-cycles, denoted as $\mathbb{Z}_{k}(|\mathcal{K}|)$, is the kernel of the boundary map $\partial_{k}$:
\begin{align*} 
    \mathbb{Z}_{k}(|\mathcal{K}|) &= \text{kernel}\Big\{ \partial_{k}: \mathbb{C}_{k}(|\mathcal{K}|) \rightarrow \mathbb{C}_{k-1}(|\mathcal{K}|)\Big\},\\
    &= \Big\{\sigma_k \in \mathbb{C}_{k}(|\mathcal{K}|) : \partial_{k}(\sigma_k) = 0 \Big\}.
\end{align*}
The elements of $\mathbb{Z}_{k}(|\mathcal{K}|)$ are called $k$-cycles and represent closed $k$-chains with no boundary. 

The space of $k$-boundaries, denoted as $\mathbb{B}_{k}(|\mathcal{K}|)$, is the image of the boundary map $\partial_{k+1}$:
\begin{align*}
    \mathbb{B}_{k}(|\mathcal{K}|) &= \text{image}\Big\{ \partial_{k+1}: \mathbb{C}_{k+1}(|\mathcal{K}|) \rightarrow \mathbb{C}_{k}(|\mathcal{K}|) \Big\},\\
   &= \Big\{ \sigma_k \in \mathbb{C}_{k}(|\mathcal{K}|) : \exists ~\tau_{k+1} \in \mathbb{C}_{k+1}(|\mathcal{K}|)~ \text{such that},~ \partial_{k+1}(\tau_{k+1}) = \sigma_k \Big\}.
\end{align*}
Elements of $\mathbb{B}_{k}(|\mathcal{K}|)$ are called $k$-boundaries, which are $k$-chains arising as boundaries of $(k+1)$-chains. The inclusion $\mathbb{B}_{k}(|\mathcal{K}|) \subseteq \mathbb{Z}_{k}(|\mathcal{K}|)$ follows from $\partial_{k}\circ \partial_{k-1}=0$ represents every boundary is necessarily a cycle. However, not all cycles are boundaries.
\paragraph{\textbf{Homology Group:}}
  The distinction between cycles and boundaries directly leads to homology. It reveals the existence of $k$-dimensional voids that are enclosed by cycles but do not arise as boundaries of higher-dimensional simplices. The $k$-th homology group $\mathbb{H}_{k}(|\mathcal{K}|)$ is defined by the quotient group: 
  \begin{equation*}
  \mathbb{H}_{k}(|\mathcal{K}|)={\mathbb{Z}_{k}(|\mathcal{K}|)}/ {\mathbb{B}_{k}(|\mathcal{K}|)}.   
  \end{equation*}
 captures equivalence classes of 
$k$-dimensional cycles that are not boundaries of $(k+1)$-chains. This homology group reflects the presence of nontrivial $k$-dimensional holes within a simplicial complex. The $k$-th Betti number
$\beta_{k} = \text{dim} \big(\mathbb{H}_{k}(|\mathcal{K}|)\big)$ quantifies the number of unique $k$-dimensional holes, and the sequence of Betti numbers $\{\beta_0, \beta_1, \beta_2, \cdots \}$ provides a summary of the topological features of the population.
\subsection{Persistent Homology and Persistence Diagram}
Topological features in data depend on the scale parameter $\varepsilon$ (see \textcite{carlsson2009topology}, section 2.3). Instead of studying topology at a fixed scale, persistent homology captures how topological structures change by varying $\varepsilon$. To capture this multi-scale topological behaviour of these features, we consider a filtration of a simplicial complex, which serves as the backbone of persistent homology.  This concept allows us to track the birth and death of topological features across multiple scales.
\subsubsection{Filtration}
A filtration of a geometric simplicial complex $|\mathcal{K}|$ is a nested sequence of geometric simplicial sub-complexes $\big\{|\mathcal{K}|_{\varepsilon}\big\}_{\varepsilon \in \mathbb{I}}$ indexed by a totally ordered set $ \mathbb{I} \subset \mathbb{R}$ such that:
\begin{enumerate}[label=(\arabic*)]
\item $ |\mathcal{K}|=\bigcup \limits_{\varepsilon \in \mathbb{I}} |\mathcal{K}|_{\varepsilon}.$
\item $|\mathcal{K}|_{\varepsilon_{i}}\subseteq |\mathcal{K}|_{\varepsilon_{j}}$ holds for all $\varepsilon_{i},\varepsilon_{j} \in \mathbb{I}$ satisfying $\varepsilon_{i} \leq \varepsilon_{j}.$
\end{enumerate}
The filtration of $|\mathcal{K}|$ is denoted as $\mathscr{F}(|\mathcal{K}|)$, and the pair $\Big(|\mathcal{K}|, \big\{|\mathcal{K}|_{\varepsilon}\big\}_{\varepsilon \in \mathbb{I}}\Big) $ is called the filtered simplicial complex. The goal is to study the evolution of topological features as filtration progresses.
\subsubsection{Persistent Homology}
 For each dimension $k \geq 0$, persistence homology studies the sequence of homology groups connected by linear maps $f_{i,j}$:
    \begin{equation*}
     \mathbb{H}_{k}(|\mathcal{K}|_{\varepsilon_{0}}) \xrightarrow{f_{0,1}} \mathbb{H}_{k}(|\mathcal{K}|_{\varepsilon_{1}})\xrightarrow{f_{1,2}}\mathbb{H}_{k}(|\mathcal{K}|_{\varepsilon_{2}})\xrightarrow{f_{2,3}}\cdots\xrightarrow{f_{n-1,n}}\mathbb{H}_{k}(|\mathcal{K}|_{\varepsilon_{n}}),
     \end{equation*}
where $f_{i,j}: \mathbb{H}_{k}(|\mathcal{K}|_{\varepsilon_{i}})\to\mathbb{H}_{k}(|\mathcal{K}|_{\varepsilon_{j}})$ is induced by the inclusion $|\mathcal{K}|_{\varepsilon_{i}}\hookrightarrow|\mathcal{K}|_{\varepsilon_{j}}$ for $\varepsilon_i \leq \varepsilon_j$. These maps track which topological features remain identifiable as $\varepsilon$ increases. The $k$-th persistence homology group between $\mathbb{H}_{k}(|\mathcal{K}|_{\varepsilon_{i}})$ and $\mathbb{H}_{k}(|\mathcal{K}|_{\varepsilon_{j}})$ is defined as:
    \begin{equation*}
        \mathbb{H}_{k}^{i,j}= \text{image}\Big\{ f_{i,j}:\mathbb{H}_{k}(|\mathcal{K}|_{\varepsilon_{i}})\rightarrow \mathbb{H}_{k}(|\mathcal{K}|_{\varepsilon_{j}})\Big\}.
    \end{equation*}
This group captures the $k$-dimensional topological features that persist across the interval $[\varepsilon_i,\varepsilon_j]$. Thus, Persistent homology provides a multiscale summary of the topological structure in data, which is essential for robust feature detection.
\subsubsection{Persistence Diagram}
For a filtered simplicial complex $\mathscr{F}(|\mathcal{K}|)=\big\{|\mathcal{K}|_{\varepsilon}\big\}_{\varepsilon \in \mathbb{I}}$, the evolution of topological features across scales is described as follows:

    \textbf{Birth: }A topological feature $ x_i^{(k)} \in \mathbb{H}_{k}\big(|\mathcal{K}|_{\varepsilon_{i}}\big)$ is said to be  born at scale $\varepsilon_{i} \in \mathbb{I}$ if, 
    $$x_i^{(k)} \notin \mathbb{H}_{k}\big(|\mathcal{K}|_{\varepsilon_{i}-\delta}\big), ~\forall~ \delta >0.$$
    
    \textbf{Death: }A topological feature $ y_{i,j}^{(k)} \in \mathbb{H}_{k}\big(|\mathcal{K}|_{\varepsilon_{i}}\big)$ born at scale $\varepsilon_{i}$, is said to die at scale $\varepsilon_{j} \in \mathbb{I}$ if,
    $$y_{i,j}^{(k)} \in \mathbb{H}_{k}\big(|\mathcal{K}|_{\varepsilon_{j}-\delta}\big),~ \forall~ \delta >0 \quad \text{satisfying}\quad \varepsilon_{i}\leq \varepsilon_{j}-\delta\quad \text{and} \quad y_{i,j}^{(k)} \notin \mathbb{H}_{k}\big(|\mathcal{K}|_{\varepsilon_{j}}\big).$$
    A feature will also die if it merges with a feature born earlier in the filtration.
    
    \textbf{Persistence Interval:} Each feature is associated with a persistence interval $[ \varepsilon_{i},\varepsilon_{j}]$, which indicates that it was born at $\varepsilon_{i}$ and died at $\varepsilon_{j}$.
    
For homology dimension $k \geq 0$, the $k$-dimensional persistence diagram $\mathrm{Dgm}_k\big(\mathscr{F} (|\mathcal{K}|)\big)$ of a filtered simplicial complex $\mathscr{F}(|\mathcal{K}|)$ is a multiset of points $\big(\varepsilon_{b_{i}}, \varepsilon_{d_{i}}\big) \in \Delta \subset \mathbb{R}^2$, where $\Delta = \big\{\big(\varepsilon_{b_{i}}, \varepsilon_{d_{i}}\big):0 \leq \varepsilon_{b_{i}} \leq \varepsilon_{d_{i}} \leq \infty \big\}$. Each point $(\varepsilon_{b_{i}},\varepsilon_{d_{i}})$ corresponds to a topological feature born at scale $\varepsilon_{b_{i}}$ and dies at scale $\varepsilon_{d_{i}}$. 
\paragraph{\textbf{Bottleneck Distance: }} To compare the topological structures of two datasets $\mathbb{X}_n=\{X_1, X_2,\cdots, X_n\}$ and $\mathbb{Y}_n=\{Y_1, Y_2,\cdots, Y_n\}$ with empirical distributions $\mathbb{P}_n$ and $\mathbb{Q}_n$, respectively, one applies the Vietoris–Rips filtration over their corresponding empirical supports $\textbf{Supp}(\mathbb{P}_n)$ and $\textbf{Supp}(\mathbb{Q}_n)$, respectively. This yields persistence diagrams $\mathrm{Dgm}_{k}\big(\mathscr{F}(\textbf{Supp}(\mathbb{P}_n))\big)$ and $\mathrm{Dgm}_{k}\big(\mathscr{F}(\textbf{Supp}(\mathbb{Q}_n))\big)$, respectively. The bottleneck distance between these diagrams in dimension $k$ is defined as:
\begin{equation*}
    W_{\infty}\Big(\mathrm{Dgm}_{k}\big(\mathscr{F}(\textbf{Supp}(\mathbb{P}_n))\big),\mathrm{Dgm}_{k}\big(\mathscr{F}(\textbf{Supp}(\mathbb{Q}_n))\big)\Big)=\inf_{\gamma_{k}^{}}~ \sup_{x\in \mathrm{Dgm}_{k}(\mathscr{F})}~||x-\gamma_{k}(x)||_{\infty} ,
\end{equation*}
where $\gamma_{k}:\mathrm{Dgm}_{k}\big(\mathscr{F}(\textbf{Supp}(\mathbb{P}_n))\big) \rightarrow ~\mathrm{Dgm}_{k}\big(\mathscr{F}(\textbf{Supp}(\mathbb{Q}_n))\big)$ is a bijection.
\subsection{Definitions}\label{sec(3.3)}
\paragraph{\textbf{Hausdorff Distance:}} To measure the closeness between two non-empty compact subsets of a metric space, we adopt the Hausdorff distance. Let $(\mathcal{M},d)$ be a metric space. For any two non-empty compact sets $A,B \subset \mathcal{M}$, the Hausdorff distance between $A$ and $B$ is defined as
\begin{equation*}\label{Def 1}
d_H(A, B) := \max\bigg\{\sup_{a\in A}\inf_{b\in B}~d(a, b), ~\sup_{b\in B} \inf_{a\in A} ~d(a, b)\bigg\}.
\end{equation*}
\paragraph{\textbf{$\varepsilon$-dense set:}} Let $(\mathcal{M},d)$ be a metric space and $\mathrm{A}$ be a non-empty subset of $\mathcal{M}$. For any $\varepsilon >0$, a finite subset $\mathrm{D}=\{ u_{i_1},\cdots, u_{i_p}\} \subseteq \mathcal{M}$ is said to be an $\varepsilon$-dense set of $\mathrm{A}$ if 
\begin{equation*}
    \sup_{a \in \mathrm{A}} d(a,\mathrm{D}):= \sup_{a \in \mathrm{A}} ~\inf_{u \in \mathrm{D}} d(a,u)< \varepsilon.
\end{equation*}
\paragraph{\textbf{Covering Number:}} Let $(\mathcal{M},d)$ be a metric space and $\mathrm{A}$ be a non-empty subset of $\mathcal{M}$.
The $\varepsilon$-covering number of $\mathrm{A}$ is the minimal cardinality of an $\varepsilon$-dense set of $\mathrm{A}$, and it is denoted by $\mathbf{N}(\varepsilon,\mathrm{A},d) \equiv \mathbf{N}$. In notation,
\begin{equation*}\label{Def 2}
    \mathbf{N}(\varepsilon,\mathrm{A},d)=\min\bigg\{p\in \mathbb{N}: \exists~ \{ u_{i_1},\cdots, u_{i_p}\} \subseteq \mathcal{M},~ \mathrm{A} \subseteq \bigcup_{j=1}^{p} B_{\varepsilon}(u_{i_j})\bigg\}.
\end{equation*}
\paragraph{\textbf{Packing Number:}} Let $(\mathcal{M},d)$ be a metric space and $\mathrm{A}$ be a non-empty subset of $\mathcal{M}$. The $\varepsilon$-packing number of $\mathrm{A}$, denoted by $\mathbf{N}'(\varepsilon, \mathrm{A},d) \equiv \mathbf{N}'$, is defined as:
\begin{equation*}\label{Def 3}
    \mathbf{N}'(\varepsilon,\mathrm{A},d)= \max\Big\{k \in \mathbb{N} \colon \exists \{x_{i_1}, \dots, x_{i_k}\} \subseteq A \text{ such that } d(x_{i_j}, x_{i_l}) > \varepsilon \text{ for all } j \neq l\Big\}.
\end{equation*}
\paragraph{\textbf{Packing Number Bound:}} Let $(\mathcal{M},d)$ be a metric space and $\mathrm{A}$ be a nonempty subset of $\mathcal{M}$. Suppose that $\mu $ is a measure on $\mathcal{M}$ satisfying the $(a,b)$-standard assumption (see \hyperref[section 2.1]{\textcolor{blue}{Section \ref{section 2.1}}}). Then for any $\varepsilon>0$, the $\varepsilon$-packing number of $\mathrm{A}$ satisfies the following upper bound:
\begin{equation}\label{eq:2.2}
    \mathbf{N}':=\mathbf{N}'(\varepsilon,\mathrm{A},d) \leq \frac{\mu(\mathrm{A})2^b}{a \varepsilon^b}.
\end{equation}
\paragraph{\textbf{Coupling of Probability Measures:}}  Let $\mu$ and $\nu$ be two probability measures defined on measurable spaces $(\Omega_1,\mathcal{F}_1)$ and $(\Omega_2,\mathcal{F}_2)$ respectively. A coupling of $\mu$ and $\nu$ is a joint probability measure $\pi$ on the product space $(\Omega_1 \times \Omega_2, \mathcal{F}_1 \times \mathcal{F}_2)$ such that: 
$$\pi(A \times \Omega_2)=\mu(A) ~\text{and} ~\pi(\Omega_1 \times B)=\nu(B),\quad \text{for all}~ A \in \mathcal{F}_1, ~B\in \mathcal{F}_2.$$
   That is, the marginals of $\pi$ are $\mu$ and $\nu$, respectively. The set of all couplings of $\mu$ and $\nu$ is denoted by $\mathscr{C}(\mu,\nu)$ and is defined as $\mathscr{C}(\mu,\nu):=\big\{\pi \in \mathbb{P}(\Omega_1 \times \Omega_2 ) :\pi(A \times \Omega_2)=\mu(A) ~\text{and} ~\pi(\Omega_1 \times B)=\nu(B),~ \text{for all}~ A \in \mathcal{F}_1, ~B\in \mathcal{F}_2 \big\}$.
   
An equivalent formulation is given in terms of random variables. Let $X \sim \mu$ and $Y \sim \nu$ be two random variables defined on measurable spaces $(\Omega_1,\mathcal{F}_1)$ and $(\Omega_2,\mathcal{F}_2)$ respectively. A coupling of $X$ and $Y$ is a pair of random variables $(\tilde{X},\tilde{Y}):(\Omega, \mathcal{F},\mathbb{P}) \to (\Omega_1 \times \Omega_2, \mathcal{F}_1\times \mathcal{F}_2)$ defined on a common probability space $(\Omega, \mathcal{F},\mathbb{P})$ such that $\tilde{X}:\Omega \to \Omega_1$ is $\mathcal{F}/\mathcal{F}_1$ measurable with law $\mathbb{P}\circ {\tilde{X}}^{-1}=\mu$ and $\tilde{Y}:\Omega \to \Omega_2$ is $\mathcal{F}/\mathcal{F}_2$ measurable with law $\mathbb{P}\circ {\tilde{Y}}^{-1}=\nu$, and the joint distribution $\mathbb{P}\circ {(\tilde{X},\tilde{Y})}^{-1}$ defines a coupling $\pi \in \mathscr{C}(\mu,\nu) $.
\paragraph{\textbf{Maximal coupling:}} A maximal coupling of two probability measures $\mu$ and $\nu$ defined on the same measurable space $(\Omega,\mathcal{F})$ is a coupling $\pi \in \mathscr{C}(\mu, \nu)$ whose associated random variables
$(X,Y)\sim \pi$ with $X \sim \mu$ and $Y \sim \nu$, maximize $\mathbb{P}(X = Y)$ among all such couplings. In other words, $\pi$ is a coupling that achieves
$$\sup\big\{ \mathbb{P}(X = Y): (X, Y) \sim \pi,\ \pi \in \mathscr{C}(\mu, \nu) \big\}.$$
\section{Methodology}\label{section 3}
We here describe the components of our robust framework sequentially. \hyperref[section 3.1]{\textcolor{blue}{Section \ref{section 3.1}}} introduces the sampling model and the associated empirical measure. \hyperref[section 3.2]{\textcolor{blue}{Section \ref{section 3.2}}} describes the asymmetric trimming procedure based on average pairwise distances. In \hyperref[section 3.3]{\textcolor{blue}{Section \ref{section 3.3}}}, we construct a Vietoris-Rips filtration on the trimmed empirical support. Finally, \hyperref[section 3.4]{\textcolor{blue}{Section \ref{section 3.4}}} discusses the construction of persistence diagrams under trimming. 
\subsection{Sampling Model and Empirical Measure }\label{section 3.1}
Suppose we observe a random sample $\mathbb{X}_n = \{X_{1},\cdots, X_{n}\}$, drawn independently from an unknown probability measure $\mathbb{P} \in \mathscr{P}_{a,b}(\mathcal{M})$, where the class $\mathscr{P}_{a,b}(\mathcal{M})$ is defined in \hyperref[section 2.1]{\textcolor{blue}{Section \ref{section 2.1}}}. The empirical measure associated with this sample is defined as $ \mathbb{P}_{n} := \frac{1}{n} \sum_{i=1}^{n}\delta_{X_i}$, where $\delta_{X_i}$ denotes the Dirac delta measure at $X_i$, and the support of $\mathbb{P}_{n}$ is denoted as \textbf{Supp}($\mathbb{P}_n$).
\subsection{Asymmetric Trimming based on Average Pairwise Distances}\label{section 3.2}
To make topological inference more robust to outliers and heavy-tailed contamination, we construct a trimmed version of the empirical measure $\mathbb{P}_{n}$. In this procedure, we exclude observations based on the largest as well as the smallest average pairwise distances relative to the rest of the sample. For each $X_i \in \mathbb{X}_n$, we define its average pairwise distance as
\begin{equation}\label{eq:3.1}
    \bar{D}_n(X_i) =\frac{1}{n-1} \sum_{\substack{j=1 \\ j \neq i}}^{n} d(X_i,X_j),~ i=1(1)n,
\end{equation}
where $d(\cdot,\cdot)$ denotes the metric on the ambient space. Note that throughout the article, ``distance" is induced by the aforementioned metric $d(., .)$ unless mentioned otherwise. We denote this collection of average pairwise distances as $\bar{\mathbb{D}}_n :=\big\{\bar{D}_n(X_1),~\bar{D}_n(X_2),~\cdots,~\bar{D}_n(X_n)\big\}$, and let $\bar{D}_{(1)} \leq \bar{D}_{(2)} \leq \cdots \leq \bar{D}_{(n)}$ denote the corresponding order statistics, where $\bar{D}_{(i)} $ is the $i^{th}$ smallest observation in the random sample $\bar{\mathbb{D}}_n$. Afterwards, for fixed trimming proportions $\alpha_1, \alpha_2 \in [0,\frac{1}{2})$, we discard the $\lfloor{\alpha_1 n}\rfloor$ largest and $\lfloor{\alpha_2 n}\rfloor$ smallest average pairwise distances from the collection $\bar{\mathbb{D}}_n$, thereby identifying and excluding the corresponding $(\lfloor{\alpha_1 n}\rfloor+\lfloor{\alpha_2 n}\rfloor)$ sample points from $\mathbb{X}_n$. Thus, the resulting trimmed dataset comprises the remaining $\left(n - \lfloor{\alpha_1 n}\rfloor - \lfloor{\alpha_2 n}\rfloor \right)$ observations and is defined as:
\begin{equation}\label{eq:3.2}
    \mathbb{X}_{n}^{\alpha_1,\alpha_2} := \{X_{i}\in \mathbb{X}_{n}: \bar{D}_{\left(\lfloor{\alpha_2 n}\rfloor+1 \right)}\leq \bar{D}_n(X_i)\leq \bar{D}_{\left(n-\lfloor{\alpha_1 n}\rfloor \right)}\},
\end{equation}
where $\bar{D}_{(i)}$ denotes the $i$-th order statistics of $\bar{\mathbb{D}}_{n}$. The corresponding trimmed empirical measure associated with the trimmed dataset $\mathbb{X}_n^{\alpha_1,\alpha_2}$ is then defined as
\begin{equation} \label{eq:3.3}
    \mathbb{P}_n^{\alpha_1,\alpha_2}:=\frac{1}{(n-\lfloor{\alpha_1 n}\rfloor - \lfloor{\alpha_2 n}\rfloor)} \sum_{i=1}^{n}\delta_{X_i} \cdot \mathbbm{1}_{\big\{\bar{D}_{(\lfloor{\alpha_2 n}\rfloor+1)}\leq \bar{D}_n(X_i)\leq \bar{D}_{(n-\lfloor{\alpha_1 n}\rfloor)}\big\}}.
\end{equation}
 \paragraph{\textbf{Population Version of the trimmed measure:}} For $x\in$ \textbf{Supp($\mathbb{P}$)}, the average pairwise distance to any random point $Y \sim \mathbb{P}$ is defined as 
\begin{equation*}
     \bar{D}(x)=E_Y[d(x,Y)]=\int_{\text{\textbf{Supp($\mathbb{P}$)}}} d(x,y) ~d\mathbb{P}(y).
\end{equation*} 
For fixed trimming proportions $\alpha_1, \alpha_2 \in [0,\frac{1}{2})$ with $\alpha_1+\alpha_2 < 1$, the upper and lower trimming thresholds are defined as
$\bar{D}_{(1-\alpha_1)} = \inf \left\{ t \in \mathbb{R} : \mathbb{P}\left( \bar{D}(x) \leq t \right) \geq 1 - \alpha_1 \right\}$ and $\bar{D}_{(\alpha_2)} = \inf \left\{ t \in \mathbb{R} : \mathbb{P}\left( \bar{D}(x) \leq t \right) \geq \alpha_2 \right\}$ respectively. The population-level trimmed support is  given by
    $\textbf{Supp}(\mathbb{P}^{\alpha_1,\alpha_2})=\mathcal{S}^{\alpha_1,\alpha_2} = \left\{ x \in \textbf{Supp}(\mathbb{P}):\bar{D}_{(\alpha_2)}\leq \bar{D}(x) \leq \bar{D}_{(1-\alpha_1)} \right\}$,
and the corresponding trimmed population measure is defined as:
\begin{equation*}
\mathbb{P}^{\alpha_1, \alpha_2}(A):= \frac{ \mathbb{P}\big( A \cap \mathcal{S}^{\alpha_1,\alpha_2} \big) }{ \mathbb{P}\big( \mathcal{S}^{\alpha_1, \alpha_2} \big) },\quad \text{for all measurable set } A \subset \mathcal{M}.
\end{equation*}
\begin{remark}
     Observe that as $\alpha_{1}\rightarrow 0$ and $\alpha_{2}\rightarrow 0$, $\mathbb{P}^{\alpha_1, \alpha_2}(\cdot)$ converges to $\mathbb{P}_{d} (\cdot)$, where $\mathbb{P}_{d}$ is the probability distribution of the random variable $E_{Y}[d(X, Y)|X]$, where $X$ and $Y$ identically and independently distributed with probability distribution $\mathbb{P}$. For example, if $\mathbb{P}$ is the same as the standard normal distribution and $d(X, Y) = \frac{(X - Y)^{2}}{2}$, then $\mathbb{P}_{d}$ will be the same as the distribution of the random variable $M_{X}$, where $M_{X}$ follows certain shifted and scaled chi-squared distribution with 1 degree of freedom. 
\end{remark}
\subsection{Filtration Construction from Trimmed Empirical Support}\label{section 3.3}
For a given scale parameter $\varepsilon \geq 0$, the Vietoris-Rips complex $\mathbb{VR}_{\varepsilon}\Big(\textbf{Supp}\big(\mathbb{P}_{n}^{\alpha_1,\alpha_2}\big)\Big)$ constructed over the trimmed empirical support $\textbf{Supp}\big(\mathbb{P}_{n}^{\alpha_1,\alpha_2}\big)$ is defined as:\\
\resizebox{0.995\linewidth}{!}{%
\begin{minipage}{\linewidth}
\begin{align}
\mathbb{VR}_{\varepsilon}\Big(\textbf{Supp}\big(\mathbb{P}_{n}^{\alpha_1,\alpha_2}\big)\Big) = \Bigg\{ \sigma_{k}^{(i)} &= [X_{i_0}, X_{i_1}, \cdots, X_{i_k}],~ \text{$k$-simplex : } \nonumber \{ X_{i_0}, X_{i_1}, \cdots, X_{i_k}\} \subseteq\textbf{Supp}\big(\mathbb{P}_{n}^{\alpha_1,\alpha_2}\big), \nonumber \\
&\text{for } i = 0, \cdots, N, \text{ where $N$ is total number of $k$-simplices, } \nonumber \\& k = 0, \cdots, \overline{ n-\lfloor{\alpha_1 n}\rfloor-\lfloor{\alpha_2 n}\rfloor-1},~ \nonumber d(X_{i_{j}},X_{i_{l}})\leq 2\varepsilon,~ \forall j,l \in \{0,\cdots, k\}\Bigg\}.
\end{align}
\end{minipage}
}\\
As $\varepsilon$ increases, these complexes form a filtration (a nested sequence of simplicial complexes) $\mathscr{F}\big(\textbf{Supp}(\mathbb{P}_{n}^{\alpha_1,\alpha_2})\big) =\Big\{\mathbb{VR}_{\varepsilon}\big(\textbf{Supp}(\mathbb{P}_{n}^{\alpha_1,\alpha_2})\big)\Big\}_{\varepsilon \geq 0}$. This filtration captures the multiscale topological features (such as connected components, loops, voids, and higher-dimensional analogues) of the trimmed dataset. 
\subsection{Trimming-Based Construction of Persistence Diagrams}\label{section 3.4}
To capture the evolution of these multiscale topological features, we use persistent homology, a prominent tool in topological data analysis. While computing the Vietoris-Rips filtration $\mathscr{F}\big(\textbf{Supp}(\mathbb{P}_{n}^{\alpha_1,\alpha_2})\big)$ over the trimmed empirical support $\textbf{Supp}(\mathbb{P}_{n}^{\alpha_1,\alpha_2})$, persistent homology tracks the birth and death times of these features, providing a robust descriptor of the underlying support. This develops a persistence diagram, which serves as a compact summary of the underlying topological structure. For any fixed homology dimension $ k \in \mathbb{N}~ \cup \{0\}$, the corresponding persistence diagram $\widehat{\mathrm{Dgm}}_k\Big(\mathscr{F}\big(\textbf{Supp}(\mathbb{P}_{n}^{\alpha_1,\alpha_2})\big)\Big)$ is defined as,
\resizebox{1.0\linewidth}{!}{%
\begin{minipage}{\linewidth}
\begin{align*}
\widehat{\mathrm{\mathrm{Dgm}}}_k\Big(\mathscr{F}\big(\textbf{Supp}(\mathbb{P}_{n}^{\alpha_1,\alpha_2})\big)\Big)=\Bigg\{\Big( \varepsilon_{b_i}^{(k)},\varepsilon_{d_i}^{(k)}\Big)\in \bar{\mathbb{R}}^2:
    ~& 0\leq \varepsilon_{b_i}^{(k)} \leq \varepsilon_{d_i}^{(k)}\leq \infty,
    \\& \exists \text{ a k- dimensional homology class}~ x^{(k)}_{b_i,d_i}  ~\text{such that}
    \\& x^{(k)}_{b_i,d_i} \in H_k\Big(\mathbb{VR}_{\varepsilon_{b_i}^{(k)}}\big(\textbf{Supp}(\mathbb{P}_{n}^{\alpha_1,\alpha_2})\big)\Big),
    \\&x^{(k)}_{b_i,d_i} \notin H_k\Big(\mathbb{VR}_{\varepsilon_{b_i}^{(k)} -\delta_1}\big(\textbf{Supp}(\mathbb{P}_{n}^{\alpha_1,\alpha_2})\big)\Big), ~\forall \delta_{1}>0,
    \\& x^{(k)}_{b_i,d_i} \notin H_k\Big(\mathbb{VR}_{\varepsilon_{d_i}^{(k)}}\big(\textbf{Supp}(\mathbb{P}_{n}^{\alpha_1,\alpha_2})\big)\Big),
    \\& x^{(k)}_{b_i,d_i} \in H_k\Big(\mathbb{VR}_{\varepsilon_{d_i}^{(k)}-\delta_2}\big(\textbf{Supp}(\mathbb{P}_{n}^{\alpha_1,\alpha_2})\big)\Big)
    \\& ~\forall \delta_{2}>0~ \text{satisfying}~ \varepsilon_{b_i}^{(k)} \leq \varepsilon_{d_i}^{(k)}-\delta_2\Bigg\}.
\end{align*}
\end{minipage}
}\\
Here each point $\Big(\varepsilon_{b_i}^{(k)}, \varepsilon_{d_i}^{(k)}\Big)$ in this persistence diagram represents a $k$-dimensional topological feature that appears at scale $\varepsilon_{b_i}^{(k)}$ (birth time) and disappears at scale $\varepsilon_{d_i}^{(k)}$ (death time).

The population-level trimmed persistence diagram $\mathrm{\mathrm{Dgm}}_k\big(\mathscr{F}\big(\textbf{Supp}(\mathbb{P}^{\alpha_1,\alpha_2})\big)\big)$ is defined analogously, based on the trimmed support $\textbf{Supp}(\mathbb{P}^{\alpha_1,\alpha_2})$ of the underlying probability measure $\mathbb{P}^{\alpha_1,\alpha_2}$, obtained by removing the $(\alpha_1+\alpha_2)$-mass of the distribution based on average pairwise distances.\\
\resizebox{0.995\linewidth}{!}{%
\begin{minipage}{\linewidth}
\begin{align*}
   \mathrm{\mathrm{Dgm}}_k\Big(\mathscr{F}\big(\textbf{Supp}(\mathbb{P}^{\alpha_1,\alpha_2})\big)\Big)=\Bigg\{\big( \varepsilon_{b_i}^{(k)},\varepsilon_{d_i}^{(k)}\big)\in \bar{\mathbb{R}}^2:&~ 0\leq \varepsilon_{b_i}^{(k)} \leq \varepsilon_{d_i}^{(k)}\leq \infty,
    \\& \exists \text{ a k- dimensional homology class}~ x^{(k)}_{b_i,d_i}  ~\text{such that}
    \\& x^{(k)}_{b_i,d_i} \in H_k\big(\mathbb{VR}_{\varepsilon_{b_i}^{(k)}}\big(\textbf{Supp}(\mathbb{P}^{\alpha_1,\alpha_2})\big)\big), 
    \\&x^{(k)}_{b_i,d_i} \notin H_k\big(\mathbb{VR}_{\varepsilon_{b_i}^{(k)} -\delta_1}\big(\textbf{Supp}(\mathbb{P}^{\alpha_1,\alpha_2})\big)\big), ~\forall \delta_{1}>0,
    \\& x^{(k)}_{b_i,d_i} \notin H_k\big(\mathbb{VR}_{\varepsilon_{d_i}^{(k)}}\big(\textbf{Supp}(\mathbb{P}^{\alpha_1,\alpha_2})\big)\big),
    \\& x^{(k)}_{b_i,d_i} \in H_k\big(\mathbb{VR}_{\varepsilon_{d_i}^{(k)}-\delta_2}\big(\textbf{Supp}(\mathbb{P}^{\alpha_1,\alpha_2})\big)\big)
    \\& ~\forall \delta_{2}>0~ \text{satisfying}~ \varepsilon_{b_i}^{(k)} \leq \varepsilon_{d_i}^{(k)}-\delta_2\Bigg\}.
\end{align*}
\end{minipage}
}

\section{Large Sample Properties}\label{section 4}
\subsection{Stability and Uniform Convergence}
The primary objective of this work is to derive an upper bound on the expected bottleneck distance between the persistence diagrams computed from the trimmed empirical support and its population-level counterpart (pointwise in terms of $\alpha_1$ and $\alpha_2$), i.e.
\begin{equation*}
     \sup_{\mathbb{P}\in \mathscr{P}_{a,b}(\mathcal{M})} \mathbb{E}\bigg[W_{\infty}\bigg(\widehat{\mathrm{\mathrm{Dgm}}}_{k}\Big(\mathscr{F}\big(\textbf{Supp}(\mathbb{P}_{n}^{{\alpha_1,\alpha_2}})\big)\Big), \mathrm{\mathrm{Dgm}}_{k}\Big(\mathscr{F}\big(\textbf{Supp}(\mathbb{P}^{{\alpha_1,\alpha_2}})\big)\Big)\bigg) \bigg], 
\end{equation*}
under some regularity conditions on the underlying distribution $\mathbb{P} \in \mathscr{P}_{a,b}(\mathcal{M})$, as discussed in (\ref{eq:2.1}). It establishes the robustness of the proposed estimator by quantifying its deviation from the corresponding population-level persistence diagram. Our methodology is built on recent advances in empirical process theory, stability of persistence diagrams (e.g., \textcite{cohen2005stability}), and trimming-based robust statistics (e.g., \textcite{garcia2024robust}, \textcite{dhar2022trimmed}). 

\subsubsection{Assumptions}\label{Assumptions}
    The following regularity conditions are imposed on the distribution $\mathbb{P}$.
\begin{itemize}
\item[(A1)] \textbf{Geometric regularity of the Support:} The support $\textbf{Supp}(\mathbb{P})$ of the measure $\mathbb{P}$ is compact and satisfies a uniform interior ball condition at some scale $r>0$, that is, for all $x \in \textbf{Supp}(\mathbb{P})$, there exists $y \in \textbf{Supp}(\mathbb{P})$ such that $x \in \bar{B}_r(y)\subset \textbf{Supp}(\mathbb{P})$. Furthermore, its boundary $\partial$\hspace{0.05cm}\textbf{Supp($\mathbb{P}$)} is of class $C^2$ (twice continuously differentiable). \label{(A1)}  
\begin{remark}
The combination of uniform interior ball condition and $C^2$ smoothness of the boundary excludes excessively sharp regions (such as cusps) and ensures that the support \textbf{Supp($\mathbb{P}$)} possesses a well-defined geometric structure. Moreover, the compactness assumption prevents the support from being unbounded; otherwise, data could be spread out infinitely, making it difficult to infer meaningful topological features. The smoothness condition on the boundary of support ensures bounded curvature and excludes regions with arbitrarily small radius of curvature.
\end{remark}

\item[(A2)] \textbf{$(a,b)$-standard assumption:} There exist constants $a > 0, b > 0$, and $\varepsilon_0 \in (0,1)$ such that for all $x \in \textbf{Supp}(\mathbb{P)}$ and for all $0 < \varepsilon < \varepsilon_0$,
$$\mathbb{P}(B_\varepsilon(x)) \geq \min(1, a \varepsilon^b)$$
where $B_\varepsilon(x)$ denotes the open ball of radius $\varepsilon$ centered at $x$. \label{(A2)}

\begin{remark}
    The $(a,b)$-standard assumption provides a uniform lower bound on the local mass around any point in the support, which enforces a uniform non-degeneracy of mass at small scales. It underpins finite sample concentration bounds for empirical measures and Hausdorff distances between the empirical and population supports. This assumption is standard and widely adopted in the theory of support estimation.
\end{remark}

\item[(A3)] \textbf{Coupling condition for the trimmed sample:}    The trimming procedure and $\mathbb{P}\in\mathscr{P}_{a,b}(\mathcal{M})$ are such that there exists a collection of independent random elements $\mathbb{Z}_{n - m} = \{Z_{1}, \ldots, Z_{n - m}\}$), so that $$\delta_{n} : = \sum\limits_{i = 1}^{n - m}{\mathbb P} (Y_{i}\neq Z_{i}) = \mathcal{O}\left( \frac{\log(n-m)}{n-m} \right),$$ where $Y_{i}\in\mathbb{X}_{n}^{\alpha_1, \alpha_2}$, and $\mathbb{X}_{n}^{\alpha_1, \alpha_2}$ is the same as defined in (\ref{eq:3.2}).\label{(A3)}
\begin{remark}
 The condition (A3) indicates that the probability measure associated with original random sample and the trimming procedure should be such that one can replace the dependent random elements (after trimming) by a certain collection of independent random elements, and for sufficiently large $n$, this replacement has negligible impact since $\delta_{n}\rightarrow 0$ as $n\rightarrow\infty$ when $m$ is fixed or even $\displaystyle\lim_{n\rightarrow\infty}\frac{m(n)}{n}\rightarrow c\in (0, 1)$. 
\end{remark}
\end{itemize}
\subsection{Main Results} 
\hyperref[theorem 1]{\textcolor{blue}{Theorem \ref{theorem 1}}} states the convergence rate of the trimmed persistence diagram after location transformation. 
\begin{theorem}\label{theorem 1}
    Let $(\mathcal{M},d)$ be a compact metric space and $\mathbb{P} \in \mathscr{P}_{a,b}(\mathcal{M})$ be a probability measure supported on $\textbf{Supp}(\mathbb{P}) \subset \mathcal{M}$, satisfying assumptions (A1) - (A3). Let $\mathbb{X}_n = \{X_1, \cdots, X_n\}$ be an i.i.d. sample drawn from $\mathbb{P}$. For some fixed $\alpha_1,\alpha_2 \in [0,1/2)$, let $\mathbb{X}_n^{\alpha_1,\alpha_2}=\{X_{i_1}, \cdots, X_{i_{n-\lfloor{\alpha_1 n}\rfloor - \lfloor{\alpha_2 n}\rfloor}}\}$ denote the trimmed dataset obtained by removing the $ \lfloor \alpha_1 n \rfloor$ points with the largest and $ \lfloor \alpha_2 n \rfloor$ points with the smallest average pairwise distances, as defined in \hyperref[section 3.2]{\textcolor{blue}{Section \ref{section 3.2}}}. Let $\textbf{Supp}(\mathbb{P}_n^{\alpha_1, \alpha_2})$ and $\textbf{Supp}(\mathbb{P}^{\alpha_1, \alpha_2})$ denote the supports of $\mathbb{P}_n^{\alpha_1,\alpha_2}$ and $\mathbb{P}^{\alpha_1, \alpha_2}$, respectively, where $\mathbb{P}_n^{\alpha_1,\alpha_2}$ and $\mathbb{P}^{\alpha_1, \alpha_2}$ are the same as defined in \hyperref[section 3.2]{\textcolor{blue}{Section \ref{section 3.2}}}.
    Then for any fixed $k \in \mathbb{N}\cup \{0\}$, there exists a constant $C > 0$, depending only on $a$, $b$, $\alpha_1$ and $\alpha_2$ such that
\begin{equation*}
\scalebox{0.875}{%
    $\displaystyle
    \sup_{\mathbb{P}\in \mathscr{P}_{a,b}(\mathcal{M})} \mathbb{E}\bigg[W_{\infty}\bigg(\widehat{\mathrm{\mathrm{Dgm}}}_{k}\Big(\mathscr{F}\big(\textbf{Supp}(\mathbb{P}_{n}^{\alpha_1,\alpha_2})\big)\Big), \mathrm{\mathrm{Dgm}}_{k}\Big(\mathscr{F}\big(\textbf{Supp}(\mathbb{P}^{\alpha_1,\alpha_2})\big)\Big)\bigg) \bigg]
\leq C \left( \frac{\log(n - \lfloor{\alpha_1 n}\rfloor - \lfloor{\alpha_2 n}\rfloor)}{n - \lfloor{\alpha_1 n}\rfloor - \lfloor{\alpha_2 n}\rfloor} \right)^{1/b},$
}
\end{equation*}
where $\mathrm{Dgm}_{k}\big(\cdot \big)$ denotes $k$-dimensional persistence diagram, $W_\infty(\cdot,\cdot)$ denotes the bottleneck distance between two persistence diagrams, and $\mathscr{F}$ is a Vietoris-Rips filtration.
\end{theorem}

As the proof of \hyperref[theorem 1]{\textcolor{blue}{Theorem \ref{theorem 1}}} involves many non-trivial arguments, the sketch of the proof is provided in the following. The detailed proof is presented in the \hyperref[Appendix]{\textcolor{blue}{Appendix}}. 

\paragraph{Sketch of the proof:} By the stability theorem of persistence diagrams (see \cite{cohen2005stability}), for Vietoris-Rips filtrations, the bottleneck distance between the persistence diagrams built on two subsets of a metric space is bounded by the Hausdorff distance between those sets. In our case, using stability theorem, we hope to see that  
\small
\begin{equation*}
W_{\infty}\bigg(\widehat{\mathrm{\mathrm{Dgm}}}_{k}\Big(\mathscr{F}\big(\textbf{Supp}(\mathbb{P}_{n}^{\alpha_1,\alpha_2})\big)\Big), \mathrm{\mathrm{Dgm}}_{k}\Big(\mathscr{F}\big(\textbf{Supp}(\mathbb{P}^{\alpha_1,\alpha_2})\big)\Big)\bigg) \leq d_H\Big(\textbf{Supp}(\mathbb{P}_n^{\alpha_1,\alpha_2}),~ \textbf{Supp}(\mathbb{P}^{\alpha_1,\alpha_2})\Big),
\end{equation*}
\normalsize
Thus, the problem reduces to understanding how well the trimmed empirical support approximates the trimmed population support in Hausdorff distance, and in determine the rate at which this approximation improves as the effective sample size $(n- \lfloor{\alpha_1 n}\rfloor - \lfloor{\alpha_2 n}\rfloor)$ increases. In particular, it is enough to bound the expectation of this Hausdorff distance after rescaling by $\bigg(\frac{(n-\lfloor{\alpha_1 n}\rfloor -\lfloor{\alpha_2 n}\rfloor)}{\log(n- \lfloor{\alpha_1 n}\rfloor - \lfloor{\alpha_2 n}\rfloor)}\bigg)^{\frac{1}{b}}$ uniformly over $\mathbb{P} \in\mathscr{P}_{a,b}(\mathcal{M})$.

The main technical difficulty is that after trimming, the retained sample $\mathbb{X}_n^{\alpha_1,\alpha_2}$ is no longer i.i.d., since the trimming rule depends on the entire configuration through average pairwise distances. Assumption (A3) addresses this by postulating a coupling on the same or on a richer probability space between $\mathbb{X}_n^{\alpha_1,\alpha_2}$ and a collection of independent random variables $ \mathbb{Z}_{n-\lfloor{\alpha_1 n}\rfloor - \lfloor{\alpha_2 n}\rfloor}$ = $ \{Z_1,\cdots, Z_{n-\lfloor{\alpha_1 n}\rfloor - \lfloor{\alpha_2 n}\rfloor}\}$ such that $\Pr\left(\mathbb{X}_n^{\alpha_1,\alpha_2} \neq \mathbb{Z}_{n-\lfloor{\alpha_1 n}\rfloor - \lfloor{\alpha_2 n}\rfloor}\right)\leq \delta_n$ for some coupling discrepancy $\delta_n$ with  $\text{Law}(Z_j) \in \mathscr{P}_{a,b}(\mathcal{M)}$. The key implication of this construction is that the distribution of the dependent trimmed sample converges in distribution to that of the i.i.d. surrogate sample as $\delta_n \to 0$. Intuitively, with probability at least $(1-\delta_n)$, the trimmed sample behaves like an i.i.d. sample of same size. This allows us to derive concentration and covering bounds for $\mathbb{Z}_{n-\lfloor{\alpha_1 n}\rfloor - \lfloor{\alpha_2 n}\rfloor}$ and transfer them back to $\mathbb{X}_n^{\alpha_1,\alpha_2}$ at an additional cost of order $\delta_n$. This coupling technique is discussed and utilized in several works; see for instance: \textcite{berkes1979approximation}, \textcite{peligrad2001note}, \textcite{tyurin2010uniform}, \textcite{sason2013entropy}, \textcite{yu2018asymptotic}, \textcite{li2021efficient}.

For the surrogate sample $\mathbb{Z}_{n-\lfloor{\alpha_1 n}\rfloor - \lfloor{\alpha_2 n}\rfloor}$, we use an $\varepsilon$-covering argument combined with the (a,b)-standard assumption and a packing number bound to obtain an upper bound of order $\varepsilon^{-b}\big[\exp\big\{-(n-\lfloor{\alpha_1 n}\rfloor -\lfloor{\alpha_2 n}\rfloor)\min\{1, a \varepsilon^b\}\big\}\big]$. Transferring this bound back to $\mathbb{X}_n^{\alpha_1,\alpha_2}$ at an additional cost of order $\delta_n$, we obtain an inequality of the form
\begin{center}\vspace{-0.8cm}
    {\footnotesize
    \begin{equation*}
        \Pr \Big(d_H\big(\textbf{Supp}(\mathbb{P}_n^{\alpha_1,\alpha_2}),~ \textbf{Supp}(\mathbb{P}^{\alpha_1,\alpha_2})\big) > 2\varepsilon\Big) \leq \frac{ 2^b (1-\alpha_1-\alpha_2)}{a \varepsilon^b} \Big[\exp\big\{-(n-\lfloor{\alpha_1 n}\rfloor -\lfloor{\alpha_2 n}\rfloor)\min\{1, a \varepsilon^b\}\big\} +  ~\delta_n\Big].
    \end{equation*}
}
\end{center}
Next, we choose $\varepsilon$ in such a way so that the above inequality yields an integrable upper bound for the rescaled Hausdorff distance $\bigg(\frac{(n-\lfloor{\alpha_1 n}\rfloor - \lfloor{\alpha_2 n}\rfloor)}{\log(n-\lfloor{\alpha_1 n}\rfloor - \lfloor{\alpha_2 n}\rfloor )}\bigg)^{\frac{1}{b}}d_H\Big(\textbf{Supp}(\mathbb{P}_n^{\alpha_1,\alpha_2}),~ \textbf{Supp}(\mathbb{P}^{\alpha_1,\alpha_2})\Big)$ uniformly over $\mathbb{P}\in \mathscr{P}_{a,b}(\mathcal{M})$. Integrating this bound gives the desired uniform bound on the expectation of the rescaled Hausdorff distance.  A technical distinction between the cases $b>1$ and $0 < b\leq 1$ lies in how the corresponding integrals are evaluated, but this affects only the calculation, not the final convergence rate. Since the coupling discrepancy $\delta_n$ satisfies assumption (A3) and hence $\delta_n \to 0$ as $n \to \infty$, so it does not affect the stated rate. Finally, combining this expectation bound with the stability theorem of persistence homology yields the same rate of convergence for the expected bottleneck distance between the $k$-dimensional persistence diagrams built over the trimmed empirical and trimmed population supports.\hfill$\Box$

\subsubsection{One-sided Trimming-based Persistence Diagram }
Let $(\mathcal{M},d)$ be a compact metric space, and $\mathbb{P} \in \mathscr{P}_{a,b}(\mathcal{M})$ be a probability measure supported on $\textbf{Supp}(\mathbb{P}) \subset \mathcal{M}$, satisfying assumptions (A1)-(A3). Let $\mathbb{X}_n = \{X_1, \cdots, X_n\}$ be an i.i.d. sample drawn from $\mathbb{P}$. For some fixed $\alpha \in [0,1)$, let $\mathbb{X}_n^{\alpha}=\{X_{i_1}, \cdots, X_{i_{n-\lfloor{\alpha n\rfloor}}}\}$ denote the trimmed dataset obtained by removing the $\lfloor \alpha n \rfloor$ points based on the largest average pairwise distances, as defined in \hyperref[section 3.2]{\textcolor{blue}{Section \ref{section 3.2}}}. The trimmed empirical measure is then defined as
        \begin{equation*}
            \mathbb{P}_n^{\alpha}=\frac{1}{n-\lfloor{\alpha n}\rfloor}\sum_{i=1}^n \delta_{X_i}\cdot \mathbbm{1}_{\big\{\bar{D}_n(X_i) \leq \bar{D}_{(n-\lfloor{\alpha n}\rfloor)}\big\}},
        \end{equation*}
        where $\bar{D}_n(X_i)$ and $\bar{D}_{(n-\lfloor{\alpha n}\rfloor)}$ are defined in \hyperref[section 3.2]{\textcolor{blue}{Section \ref{section 3.2}}}, with support $\textbf{Supp}(\mathbb{P}_{n}^{\alpha})$.
        
The corresponding population version of the trimming rule is defined as follows: For $x\in$ \textbf{Supp($\mathbb{P}$)}, the average pairwise distance to any random point Y $\sim \mathbb{P}$ is defined as 
\begin{equation*}
     \bar{D}(x)=E_Y[d(x,Y)]=\int_{\text{\textbf{Supp($\mathbb{P}$)}}} d(x,y) ~d\mathbb{P}(y).
\end{equation*} 
For fixed trimming proportion $\alpha \in [0,1)$, the trimming threshold is defined as
$\bar{D}_{(1-\alpha)} = \inf \left\{ t \in \mathbb{R} : \mathbb{P}\left( \bar{D}(x) \leq t \right) \geq 1 - \alpha \right\}$. The population-level trimmed support is then given by
    $\textbf{Supp}(\mathbb{P}^{\alpha})=\mathcal{S}^{\alpha} = \left\{ x \in \textbf{Supp}(\mathbb{P}): \bar{D}(x) \leq \bar{D}_{(1-\alpha)} \right\}$,
and the corresponding trimmed population measure is defined as:
\begin{equation*}
\mathbb{P}^{\alpha}(A):= \frac{ \mathbb{P}\big( A \cap \mathcal{S}^{\alpha} \big) }{ \mathbb{P}\big( \mathcal{S}^{\alpha} \big) },\quad \text{for all measurable set } A \subset \mathcal{M}.
\end{equation*}

Here, \hyperref[Corollary 1.1]{\textcolor{blue}{Corollary \ref{Corollary 1.1}}} studies an analogous result to \hyperref[theorem 1]{\textcolor{blue}{Theorem \ref{theorem 1}}} for one sided trimming.

\begin{corollary}[\textbf{Convergence of Persistence Diagrams under One-sided Trimming}]
For any fixed $k \in \mathbb{N}\cup \{0\}$, there exists a constant $C_{a,b,\alpha} > 0$ such that
\small
\begin{equation*}
    \sup_{\mathbb{P}\in \mathscr{P}_{a,b}(\mathcal{M})} \mathbb{E}\bigg[W_{\infty}\bigg(\widehat{\mathrm{\mathrm{Dgm}}}_{k}\Big(\mathscr{F}\big(\textbf{Supp}(\mathbb{P}_{n}^{\alpha})\big)\Big), \mathrm{\mathrm{Dgm}}_{k}\Big(\mathscr{F}\big(\textbf{Supp}(\mathbb{P}^{\alpha})\big)\Big)\bigg) \bigg]
\leq C_{a,b,\alpha} \left( \frac{\log(n - \lfloor{\alpha n\rfloor})}{n - \lfloor{\alpha n\rfloor}} \right)^{1/b}.
\end{equation*}
\normalsize
\label{Corollary 1.1}
\end{corollary}
\begin{remark}\nonumber
    \hyperref[Corollary 1.1]{\textcolor{blue}{Corollary \ref{Corollary 1.1}}} highlights that, under contamination from heavy-tailed distributions, one-sided trimming based on the highest average pairwise distances ensures stable persistence diagrams. Moreover, the expected bottleneck distance between persistence diagrams computed from trimmed empirical supports and their population counterparts achieves the optimal convergence rate, up to logarithmic factors. 
\end{remark}
\begin{remark}
One-sided trimming is particularly useful when contamination arises through a small number of points that lie far from the main data cloud. Therefore, these extreme observations exhibit disproportionately large average pairwise distances than the rest of the sample, often coming from heavy-tailed noise or measurement errors. They can distort persistent homology by introducing spurious features in the persistence diagrams. By removing these extreme observations, one-sided trimming allows us to preserve the intrinsic geometric structure of the data and the resulting persistence diagrams provide a more stable representation of the underlying topological structure.
\end{remark}
\section{Data Driven Choice of Trimming Proportions} \label{section 5}
The stability and uniform convergence results established in the previous section are derived for fixed trimming proportions $(\alpha_1,\alpha_2)$, and these results remain valid for any admissible choice of trimming proportions. However, in reality, the trimming proportions are unknown in practice, and to overcome this issue, in this section, we describe a procedure for choosing trimming proportions in data analysis. \hyperref[Algorithm 1]{\textcolor{blue}{Algorithm \ref{Algorithm 1}}} outlines a method for selecting $(\alpha_1,\alpha_2)$ in the asymmetric trimming procedure, while \hyperref[Algorithm 2]{\textcolor{blue}{Algorithm \ref{Algorithm 2}}} describes the selection procedure of $\alpha$ for one-sided trimming case.
\begin{algorithm}[!htbp]
\small
\caption{Selection of Asymmetric Trimming proportions for Robust Persistent Homology}
 \vspace{8pt}
 \textbf{Input:}\hspace{15pt}Sample $\mathbb{X}_n = \{X_1, \dots, X_n\} \subset \mathcal{M}$ with metric $d$;\vspace{3pt}  \\ \vspace{3pt}
  \hspace*{44pt} Initial trimming levels  $\alpha_1^0, \alpha_2^0 \in [0, 1/2)$, Step size  $\Delta \alpha_1 > 0$, $\Delta \alpha_2 > 0$; \\ \vspace{3pt}
    \hspace*{44pt} Homology dimension $k \in \mathbb{N} \cup\{0\}$, Filtration parameter $\varepsilon \in \mathbb{R}^+ \cup \{0\}$; \\ \vspace{3pt}
 \hspace*{44pt} Persistence threshold  $\tau_{\min} > 0$, maximum iterations $T \in \mathbb{N} $.\\
 \textbf{Output:}  Robust persistence diagram $\mathrm{Dgm}_k^*$, trimming levels $(\alpha_1^*,~ \alpha_2^*)$. \vspace{3pt}
\begin{algorithmic}[1]
\State Initialize $\alpha_1 \gets \alpha_1^0$ ,  $\alpha_2 \gets \alpha_2^0$ , $t \gets 0$.
\vspace{3pt}
\State Initialize $\mathrm{Dgm}_k^* \gets \varnothing$.
\While{$t < T$} \vspace{3pt}
    \For{$i = 1$ to $n$}  \vspace{3pt}
    \State $\bar{D}_n(X_i) \gets \frac{1}{n-1} \sum_{j \ne i} d(X_i, X_j)$. \vspace{3pt}
    \EndFor  \vspace{3pt}
    \State Sort $\{\bar{D}_n(X_i) \}$  to obtain order statistics  $\bar{D}_{(1)} \le \cdots \le \bar{D}_{(n)}$.  \vspace{3pt}
    \State Define trimmed sample:  $\mathbb{X}_n^{\alpha_1,\alpha_2}  \gets \{ X_i \in \mathbb{X}_n: \bar{D}_{(\lfloor{\alpha_2 n}\rfloor + 1)} \leq \bar{D}_n(X_i) \le \bar{D}_{(n - \lfloor{\alpha_1 n}\rfloor)} \}. $ \vspace{3pt}
    \State Define trimmed empirical measure: $\mathbb{P}_n^{\alpha_1,\alpha_2}$. \vspace{3pt}
    \State Define  trimmed empirical support: $\textbf{Supp}(\mathbb{P}_{n}^{\alpha_1,\alpha_2})$. \vspace{3pt}
    \State Compute Vietoris-Rips Complex: $\mathbb{VR}_{\varepsilon}\big(\textbf{Supp}(\mathbb{P}_{n}^{\alpha_1,\alpha_2})\big)$, for $\varepsilon \geq 0$. \vspace{3pt}
    \State Construct Vietoris-Rips filtration: $\mathscr{F}\big(\textbf{Supp}(\mathbb{P}_{n}^{\alpha_1,\alpha_2})\big) =\Big\{\mathbb{VR}_{\varepsilon}\big(\textbf{Supp}(\mathbb{P}_{n}^{\alpha_1,\alpha_2})\big)\Big\}_{\varepsilon \geq 0}$. \vspace{3pt}
    \State Compute persistence diagram:  $\mathrm{Dgm}_k^{(t)} \gets \mathrm{Dgm}_k\Big(\mathscr{F}\big(\textbf{Supp}(\mathbb{P}_{n}^{\alpha_1,\alpha_2})\big)\Big)$.  \vspace{3pt}
    \If{$\exists$  $(\varepsilon_{b_i},~\varepsilon_{d_i}) \in \mathrm{Dgm}_k^{(t)} $ with  $(\varepsilon_{d_i}-\varepsilon_{b_i}) \ge \tau_{\min}$ } \vspace{3pt}
        \State  $\mathrm{Dgm}_k^* \gets \mathrm{Dgm}_k^{(t)}, \quad (\alpha_1^*,~ \alpha_2^*) \gets (\alpha_1, ~\alpha_2) $  
        \State \textbf{break} \vspace{3pt}
    \Else  \vspace{3pt}
       \State $\alpha_1 \gets \max(0,~ \alpha_1 - \Delta\alpha_1)$,\quad $\alpha_2 \gets \max(0, ~\alpha_2 - \Delta\alpha_2)$,\quad $t \gets t + 1$ \vspace{3pt}
    \EndIf  \vspace{3pt}
\EndWhile  \vspace{3pt}
\If{$\mathrm{Dgm}_k^* = \varnothing$}
\State $\mathrm{Dgm}_k^* \gets \mathrm{Dgm}_k^{(T)},\quad(\alpha_1^*, \alpha_2^*) \gets (\alpha_1, \alpha_2)$.
\EndIf \vspace{3pt}
\State \Return  $\mathrm{Dgm}_k^*, \quad (\alpha_1^*, \alpha_2^*)$.
\end{algorithmic}
\label{Algorithm 1}
\end{algorithm}
\begin{algorithm}
\small
\caption{Selection of One-Sided Trimming Proportions for Robust Persistent Homology}
 \vspace{8pt}
 \textbf{Input:}\hspace{15pt}Sample $\mathbb{X}_n = \{X_1, \dots, X_n\} \subset \mathcal{M}$ with metric $d$;\vspace{3pt}  \\ \vspace{3pt}
  \hspace*{44pt} Initial trimming level  $\alpha\in [0, 1)$,     Step size  $\Delta \alpha > 0$; \\ \vspace{3pt}
    \hspace*{44pt} Homology dimension $k \in \mathbb{N} \cup\{0\}$, Filtration parameter $\varepsilon \in \mathbb{R}^+ \cup \{0\}$; \\ \vspace{3pt}
    \hspace*{44pt} Persistence threshold  $\tau_{\min} > 0$, maximum iterations $T \in \mathbb{N}$.\\
 \textbf{Output:}  Robust persistence diagram $\mathrm{Dgm}_k^*$, trimming levels $\alpha^*$. \vspace{3pt}
\begin{algorithmic}[1]
\State Initialize $\alpha \gets \alpha^0$, $t \gets 0$.
\State Initialize $\mathrm{Dgm}_k^* \gets \varnothing$.
\vspace{3pt}
\While{$t < T$} \vspace{3pt}
    \For{$i = 1$ to $n$}  \vspace{3pt}
    \State $\bar{D}_n(X_i) \gets \frac{1}{n-1} \sum_{j \ne i} d(X_i, X_j)$. \vspace{3pt}
    \EndFor  \vspace{3pt}
    \State Sort $\{\bar{D}_n(X_i) \}$  to obtain order statistics  $\bar{D}_{(1)} \le \cdots \le \bar{D}_{(n)}$.  \vspace{3pt}
    \State Define trimmed sample:  $\mathbb{X}_n^{\alpha}  \gets \{ X_i \in \mathbb{X}_n: \bar{D}_n(X_i) \le \bar{D}_{(n - \lfloor{\alpha n\rfloor})} \}. $ \vspace{3pt}
    \State Define trimmed empirical measure: $\mathbb{P}_n^{\alpha}$. \vspace{3pt}
    \State Define  trimmed empirical support: $\textbf{Supp}(\mathbb{P}_{n}^{\alpha})$. \vspace{3pt}
    \State Compute Vietoris-Rips Complex: $\mathbb{VR}_{\varepsilon}\big(\textbf{Supp}(\mathbb{P}_{n}^{\alpha})\big)$, for $\varepsilon \geq 0$. \vspace{3pt}
    \State Construct Vietoris-Rips filtration: $\mathscr{F}\big(\textbf{Supp}(\mathbb{P}_{n}^{\alpha})\big) =\Big\{\mathbb{VR}_{\varepsilon}\big(\textbf{Supp}(\mathbb{P}_{n}^{\alpha})\big)\Big\}_{\varepsilon \geq 0}$. \vspace{3pt}
    \State Compute persistence diagram:  $\mathrm{Dgm}_k^{(t)} \gets \mathrm{Dgm}_k\Big(\mathscr{F}\big(\textbf{Supp}(\mathbb{P}_{n}^{\alpha})\big)\Big)$.  \vspace{3pt}
    \If{$\exists$  $(\varepsilon_{b_i},~\varepsilon_{d_i}) \in \mathrm{Dgm}_k^{(t)} $ with  $(\varepsilon_{d_i}-\varepsilon_{b_i}) \ge \tau_{\min}$ } \vspace{3pt}
        \State  $\mathrm{Dgm}_k^* \gets \mathrm{Dgm}_k^{(t)} ,\quad \alpha^* \gets \alpha $ 
        \State \textbf{break} \vspace{3pt}
    \Else  \vspace{3pt}
       \State $\alpha \gets \max(0,~ \alpha - \Delta\alpha)$, \quad $t \gets t + 1$   \vspace{3pt}
    \EndIf  \vspace{3pt}
\EndWhile  \vspace{3pt}
\If{$\mathrm{Dgm}_k^* = \varnothing$}
\State $\mathrm{Dgm}_k^* \gets \mathrm{Dgm}_k^{(T)},\quad \alpha^* \gets  \alpha$.
\EndIf \vspace{3pt}
\State \Return  $\mathrm{Dgm}_k^*, \quad \alpha^*$.
\end{algorithmic}
\label{Algorithm 2}
\end{algorithm}
\section{Simulation Studies} \label{section 6}
This section illustrates the empirical performance of our proposed asymmetric trimming method in recovering topological features of the support of a target distribution under contamination. This contamination introduces both external and internal outliers, which affect the stability of homological inference, as homology is sensitive to outliers. For example, observations residing inside a circular hole can fill the circular hole earlier and reduce its persistence, while observations lying outside the core structure may introduce misleading topological features. We assume that the data are drawn independently from a mixture distribution
\begin{equation*}
  \mathbb{P} = (1 - \lambda) \mathbb{P}_{\text{signal}} + \lambda~ \mathbb{P}_{\text{outlier}}, \quad \lambda \in \Big(0,\frac{1}{2}\Big),
\end{equation*}
where  $\mathbb{P}_{\text{signal}}$  represents the distribution of interest, and $\mathbb{P}_{\text{outlier}}$ denotes an arbitrary contaminating distribution. Our objective is to recover the persistent topological features of $\textbf{Supp}(\mathbb{P}_{\text{signal}})$ from a finite sample $ X_1, \cdots, X_n \sim \mathbb{P} $, when $\lambda \in \Big(0,\frac{1}{2}\Big)$.
\subsection{Case Study 1: Circular Structure with Clustered Contamination}
To highlight the role of trimming in recovering meaningful topological features, we begin with an example where data is sampled around the circumference of a unit circle with clustered contamination. This setup allows us to illustrate how contamination disrupts homological inference, and how trimming restores the underlying features.

To explore the robustness of topological inference under contamination, we simulate a dataset of $n=200$ observations in $\mathbb{R}^2$, consisting of a signal component and a clustered contamination component. 

    \textbf{Signal component:} The signal component $\mathbb{P}_{\text{signal}}$ consists of 120 points sampled near a unit circle centered at the origin (0,0), forming an approximation of a one-dimensional manifold with a prominent topological feature (a persistent 1-dimensional hole). For $i=1,\cdots,120$, we generate $X_i=(r_i\cos\theta_i,~r_i\sin\theta_i) \in \mathbb{R}^2$ where $\theta_i \sim \text{Uniform}~[0,2\pi]$ and $r_i \sim \text{N}(1,0.04)$ independently.  
    
     \textbf{Contamination component:} The remaining 80 points represent contamination and are sampled from a mixture of five independent bivariate Gaussian distributions, with each cluster contributing 16 points. More precisely, the contaminated observations are generated as follows: 
     \begin{center}
    {\footnotesize
  $Y^{(k)}_{1} \sim \text{N}_2\left(
\begin{pmatrix} 1.25 \\ 1.25 \end{pmatrix},~
\sigma^2 I_2
\right), \qquad
Y^{(k)}_{2} \sim \text{N}_2\left(
\begin{pmatrix} 1.25 \\ -1.25 \end{pmatrix},~
\sigma^2 I_2
\right), \qquad
Y^{(k)}_{3} \sim \text{N}_2\left(
\begin{pmatrix} -1.25 \\ 1.25 \end{pmatrix},~
\sigma^2 I_2
\right),$\vspace{0.3cm}
$Y^{(k)}_{4} \sim \text{N}_2\left(
\begin{pmatrix} -1.25 \\ -1.25 \end{pmatrix},~
\sigma^2 I_2
\right),\qquad
Y^{(k)}_{5} \sim \text{N}_2\left(
\begin{pmatrix} 0 \\ 0 \end{pmatrix},~
\sigma^2 I_2
\right),
\quad \text{with}\quad \sigma^2 = 0.0144, \quad \text{and} \quad k=1,\cdots,16.$
}
\end{center}
Four clusters are located outside the circular signal region, and therefore, they act as external contamination, while the cluster with mean vector 
{\footnotesize $  \begin{pmatrix}
      0 \\ 0
      \end{pmatrix}$ }
produces internal contamination, which may obscure the one-dimensional hole.
     \begin{remark}
         This configuration provides a challenging test case in $\mathbb{R}^2$, for evaluating the robustness of the trimming procedure since both internal and external contamination can distort the persistent homology of the underlying circular structure.
     \end{remark}
\begin{figure}[h!]
        \centering
        \includegraphics[scale=0.13]{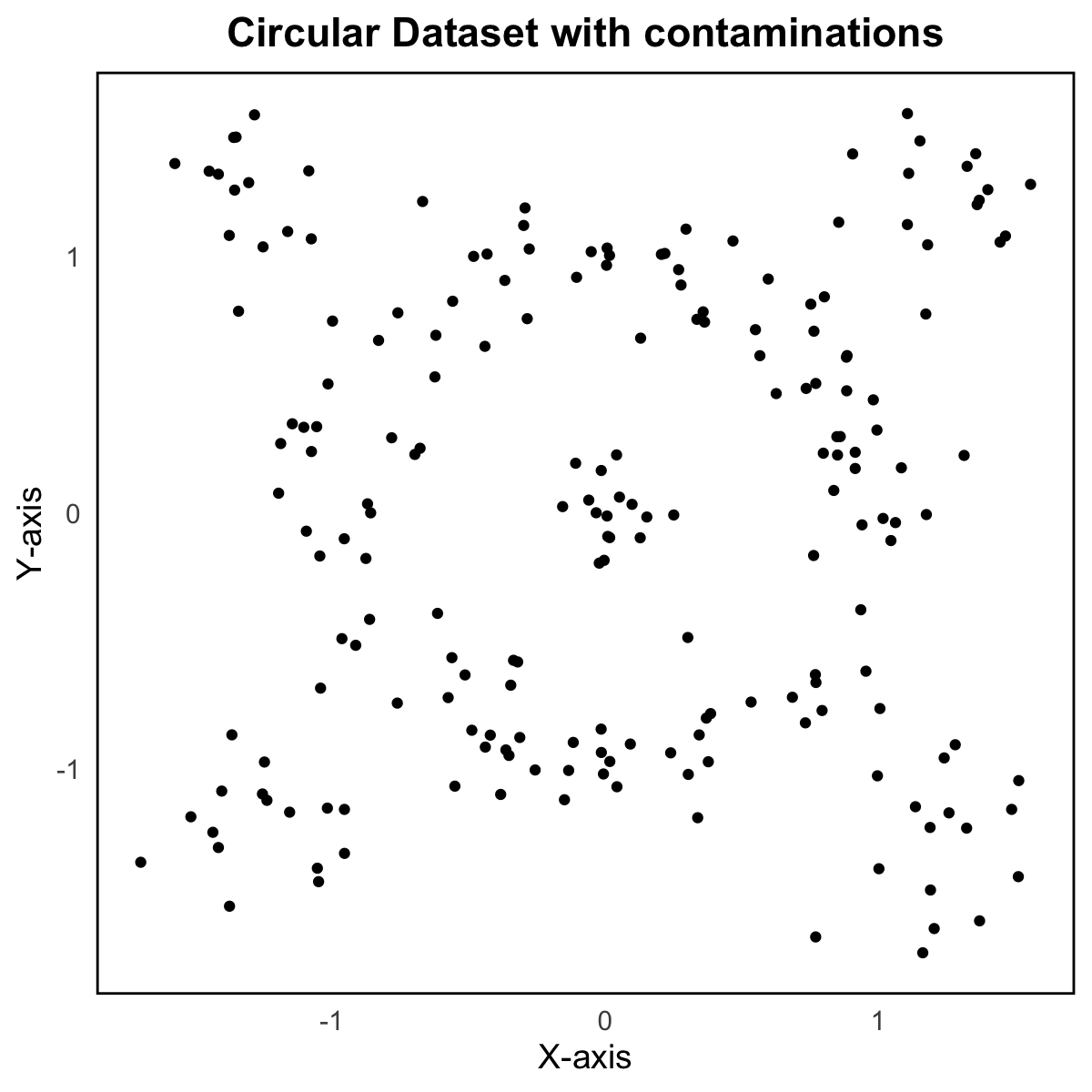}~
       \includegraphics[scale=0.13]{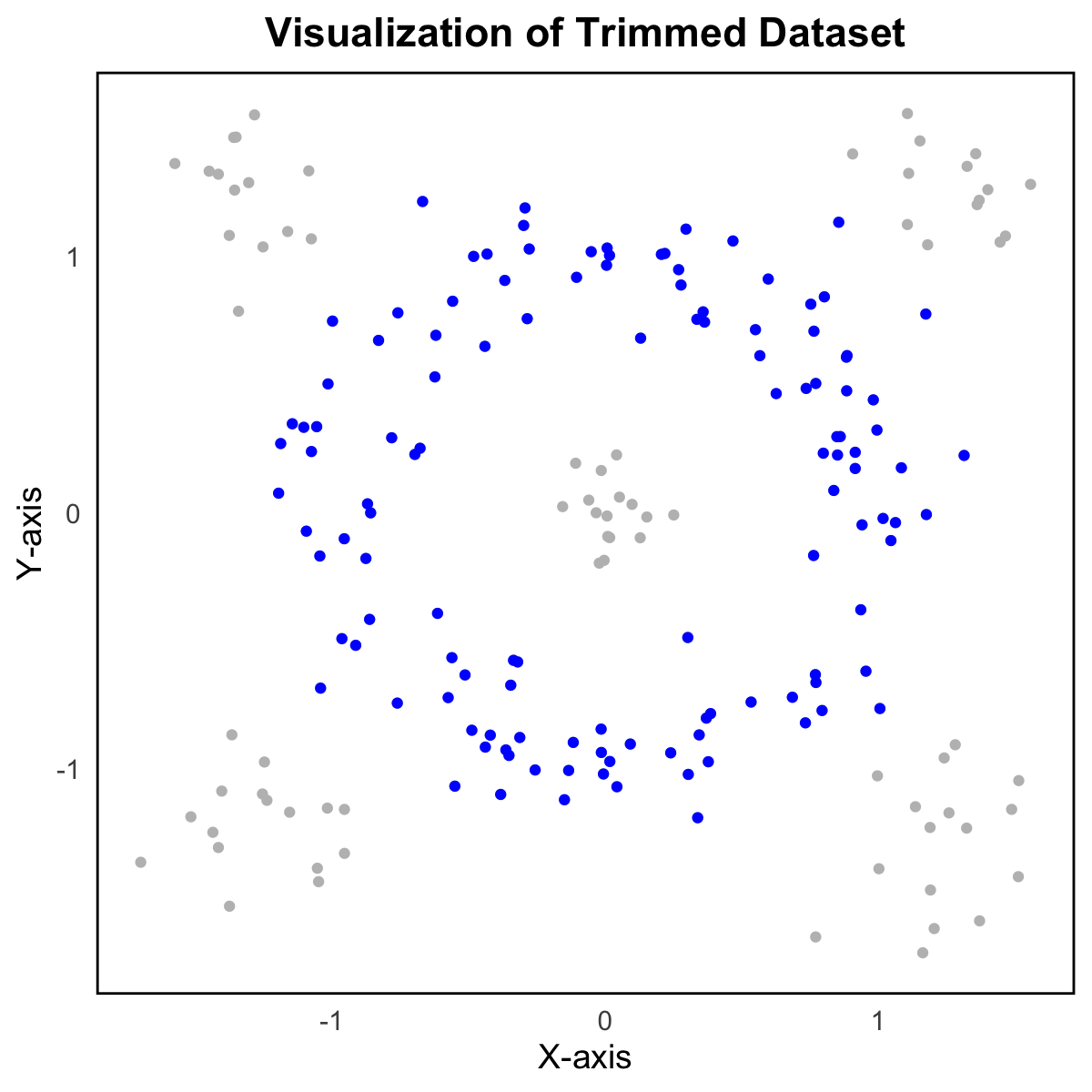}
        \caption{Comparison of the contaminated dataset (left) and the trimmed dataset (right) in the circular structure.}
        \label{Figure 6.1}
        \noindent
\end{figure}
To reduce the influence of these contaminations, we apply an asymmetric trimming strategy based on the average pairwise distances. For each point, we compute the average pairwise distance to the rest of the sample points and trim the top 30\% of points with the largest average pairwise distances and the bottom 8\% with the smallest average pairwise distances. One can observe it from \hyperref[Figure 6.1]{\textcolor{blue}{Figure \ref{Figure 6.1}}}. We then construct the Vietoris-Rips filtration both on the sample and trimmed datasets and compute their corresponding persistence diagrams. \hyperref[Figure 6.2]{\textcolor{blue}{Figure \ref{Figure 6.2}}} illustrates that the sample persistence diagram shows many short-lived one-dimensional features $H_1$, which are introduced by contamination. Here, the dominant loop appears with a persistence interval of $(0.273, 0.631)$, corresponding to a persistence length of $0.358$. Such a short interval is difficult to separate from noise and thus fails to provide information about the underlying structure.  

\hyperref[Table 6.1]{\textcolor{blue}{Table \ref{Table 6.1}}} shows that trimming changes the picture substantially by summarizing the results under various trimming proportions. With even slight changes in trimming parameters, for instance, $(\alpha_1,\alpha_2) = (0.3, 0.08)$, the dominant $H_1$ feature extends to the interval $(0.273, 1.227)$ with a persistence length of $0.954$, more than double relative to the untrimmed case. Increasing the lower trim to $\alpha_2 = 0.10$ yields an even stronger loop, with persistence length reaching $1.12$, over three times larger than the untrimmed case. In contrast, the trimmed persistence diagram reveals a single prominent loop, corresponding to the true circular structure.
\begin{figure}[h!]
        \centering
        \includegraphics[scale=0.25]{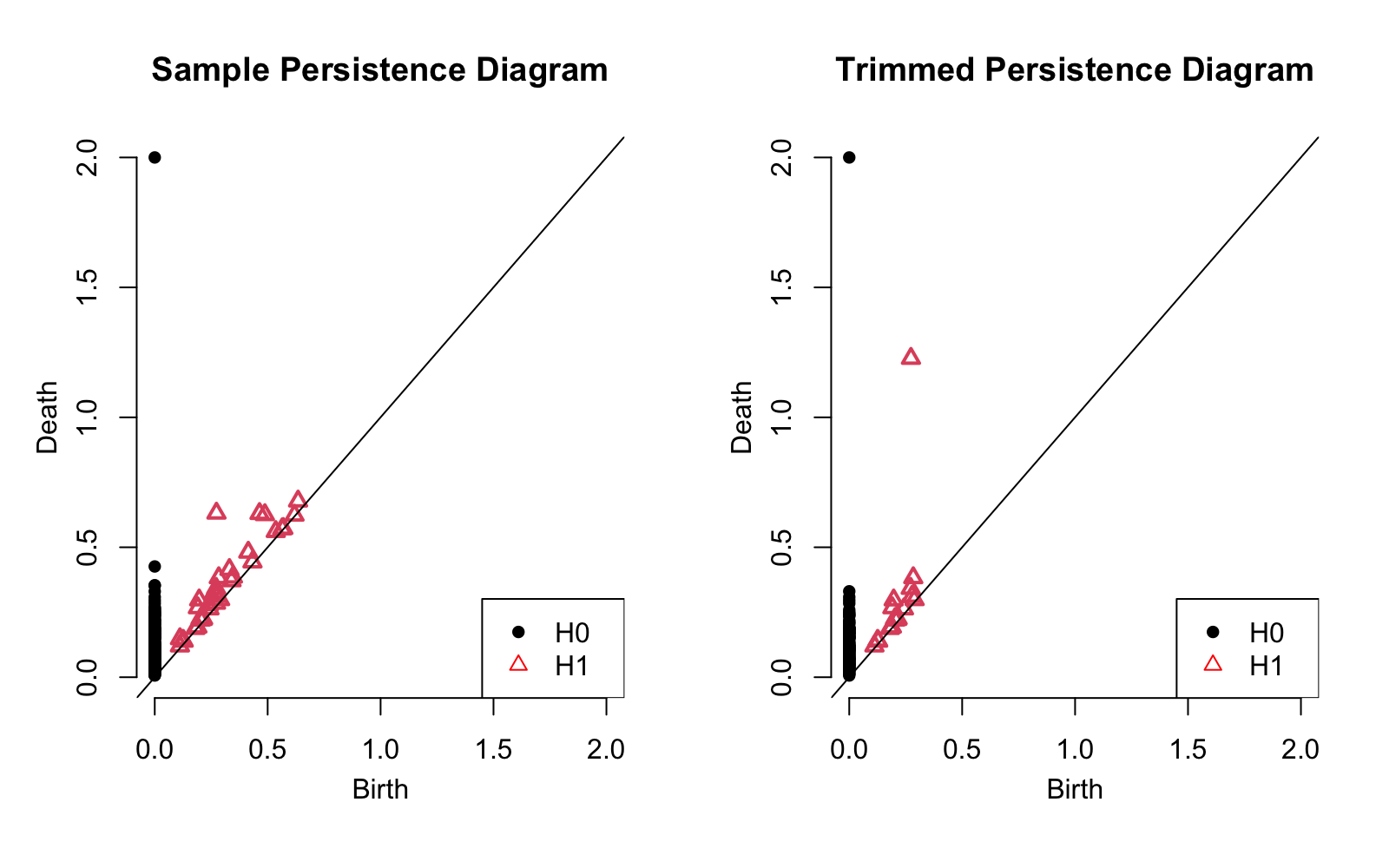}
        \caption{Persistence diagrams of the contaminated dataset (left) and the trimmed dataset (right), showing the recovery of the dominant topological feature after trimming.}
        \label{Figure 6.2}
\end{figure}

\begin{table}[h!]
\centering
\renewcommand{\thetable}{6.1}
\caption{Comparison of the highest persistence feature in $H_1$ of both the sample and the trimmed persistence diagrams of the circular structure with clustered contamination}
\footnotesize
\begin{tabular}{ccccccc}
\toprule
$n$ & \multicolumn{2}{c}{Sample Persistence Diagram} & \multicolumn{2}{c}{Trimming Proportion} & \multicolumn{2}{c}{Trimmed Persistence Diagram} \\
\cmidrule(lr){2-3} \cmidrule(lr){4-5} \cmidrule(lr){6-7}
& Persistence Interval & Length & \quad\quad $\alpha_1$ & \quad$\alpha_2$ & Persistence Interval & Length \\
\midrule
200 & (0.2732632, 0.6314332)  & 0.3581699 & \quad\quad 0.1 & \quad 0.01  & (0.2732632, 0.6314332)  & 0.3581699 \\

     &  &  & \quad\quad 0.2 & \quad 0.01  & (0.2732632, 0.6314332)  & 0.3581699 \\

     &   &  & \quad\quad 0.3 & \quad 0.01  & (0.2732632, 0.6314332)  & 0.3581699\\ \\

     &  &  & \quad\quad 0.1 & \quad 0.05  & (0.2732632, 0.6314332)  & 0.3581699\\
     
     &  &  & \quad\quad 0.2 & \quad 0.05  & (0.2732632, 0.6314332)  & 0.3581699\\
     
     &  &  & \quad\quad 0.3 & \quad 0.05  & (0.2732632, 0.6314332)  & 0.3581699\\ \\

     &  &  & \quad\quad 0.1 & \quad 0.07  & (0.2732632, 0.807147)  & 0.5338837\\
     
     &  &  & \quad\quad 0.2 & \quad 0.07  & (0.2732632, 0.807147)  & 0.5338837\\
     
     &  &  & \quad\quad 0.3 & \quad 0.07  & (0.2732632, 0.807147)  & 0.5338837\\ \\

     &  &  & \quad\quad 0.1 & \quad 0.08  & ((0.2732632, 1.227019)  & 0.9537553\\
     
     &  &  & \quad\quad 0.2 & \quad 0.08  & (0.2732632, 1.227019)  & 0.9537553\\
     
     &  &  & \quad\quad 0.3 & \quad 0.08  & (0.2732632, 1.227019)  & 0.9537553\\ \\

     &  &  & \quad\quad 0.1 & \quad 0.10  & (0.2732632, 1.397292)  & 1.124029\\
     
     &  &  & \quad\quad 0.2 & \quad 0.10  & (0.2732632, 1.397292)  & 1.124029\\
     
     &  &  & \quad\quad 0.3 & \quad 0.10  & (0.2732632, 1.397292)  & 1.124029\\ 
\bottomrule
\end{tabular}
\vspace{-0.3cm}
\end{table}\label{Table 6.1}

\hyperref[Figure 6.2]{\textcolor{blue}{Figure \ref{Figure 6.2}}} illustrates that the sample persistence diagram (left), that is, the persistence diagram drawn using sample data, exhibits various short-lived cycles introduced by contamination, whereas for $\alpha_1=0.3$ and $\alpha_2=0.08$, the trimmed persistence diagram (right) indicates the presence of a circular feature. This demonstrates that with properly chosen trimming proportions, it is possible to recover the true topological features of the support of the underlying distribution by suppressing the contamination.
\subsection{Case Study 2: Spherical Structure with Clustered Contamination}
To further evaluate the robustness of our asymmetric trimming method in higher dimensions, we consider a three-dimensional dataset where the signal distribution lies close to the surface of a unit sphere. This scenario highlights the ability of trimming to recover higher-dimensional topological features. In this case study, we analyze a three-dimensional dataset with a prominent two-dimensional cavity corresponding to the hollow interior of the sphere.

We simulate a dataset of size $n=400$, consisting of 265 signal points and 135 contaminating points in $\mathbb{R}^3$. 

    \textbf{Signal component:} Each observation of the signal component $\mathbb{P}_{\text{signal}}$ is generated by $\theta_i \sim \text{Uniform}~[0,\pi],~ \phi_i \sim \text{Uniform}~[0,2\pi]$, and $r_i\sim \text{N}(1,0.01)$ independently for $i=1,\cdots,265$ and mapped to Cartesian coordinates through $X_i=(r_i \sin \theta_i\cos\phi_i, ~r_i\sin\theta_i \sin\phi_i,~r_i\cos\theta_i)$, that concentrates around a two-dimensional spherical surface.
    
   \textbf{Contamination component:} The contamination component comprises 135 points from nine independent trivariate Gaussian distributions, forming nine independent clusters with each cluster containing 15 data points. Specially for $k=1,\cdots,15$, the contaminated points are drawn as 
     \begin{center}
    {\footnotesize
  $Y^{(k)}_{1} \sim \text{N}_3\left(
\begin{pmatrix} 1.01 \\ 1.01 \\ 1.01 \end{pmatrix},~
\sigma^2 I_3
\right), \qquad
Y^{(k)}_{2} \sim \text{N}_3\left(
\begin{pmatrix} 1.01 \\ 1.01 \\ -1.01\end{pmatrix},~
\sigma^2 I_3
\right), \qquad
Y^{(k)}_{3} \sim \text{N}_3\left(
\begin{pmatrix} 1.01 \\ -1.01 \\ 1.01 \end{pmatrix},~
\sigma^2 I_3
\right),$\vspace{0.3cm}
  $Y^{(k)}_{4} \sim \text{N}_3\left(
\begin{pmatrix} 1.01 \\ -1.01 \\ -1.01 \end{pmatrix},~
\sigma^2 I_3
\right), \qquad
Y^{(k)}_{5} \sim \text{N}_3\left(
\begin{pmatrix} -1.01 \\ 1.01 \\ 1.01\end{pmatrix},~
\sigma^2 I_3
\right), \qquad
Y^{(k)}_{6} \sim \text{N}_3\left(
\begin{pmatrix} -1.01 \\ 1.01 \\ -1.01 \end{pmatrix},~
\sigma^2 I_3
\right),$\vspace{0.3cm}
  $Y^{(k)}_{7} \sim \text{N}_3\left(
\begin{pmatrix} -1.01 \\ -1.01 \\ 1.01 \end{pmatrix},~
\sigma^2 I_3
\right), \qquad
Y^{(k)}_{8} \sim \text{N}_3\left(
\begin{pmatrix} -1.01 \\ -1.01 \\ -1.01\end{pmatrix},~
\sigma^2 I_3
\right), \qquad
Y^{(k)}_{9} \sim \text{N}_3\left(
\begin{pmatrix} 0 \\ 0 \\ 0 \end{pmatrix},~
\sigma^2 I_3
\right),$
}
\end{center}
with $\sigma^2 = 0.04$. Eight clusters lie outside the spherical support and act as external contamination, whereas the cluster with center at {\footnotesize $  \begin{pmatrix}
      0 \\ 0 \\ 0
      \end{pmatrix}$ } produces internal contamination that can fill the void.

\begin{figure}[h!]
        \centering
        \includegraphics[scale=0.25]{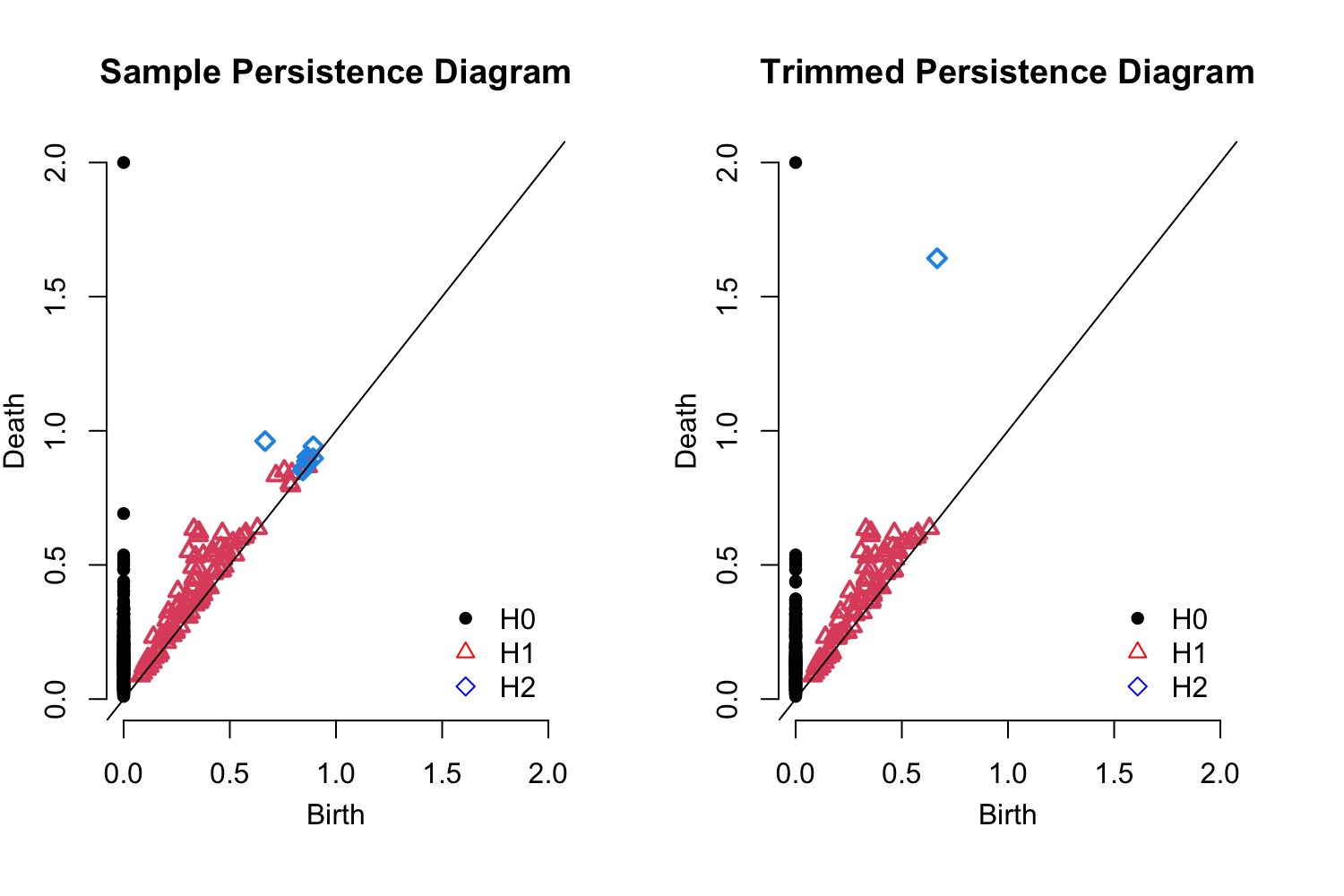}
        \caption{Persistence diagrams of the contaminated dataset (left) and the trimmed dataset (right), showing the recovery of the dominant $H_2$ topological feature after trimming.}
        \label{Figure 6.3}
\end{figure}

To bring down the influence of these contaminations, we apply our proposed asymmetric trimming procedure and then construct the Vietoris-Rips filtration both on the sample and trimmed datasets to obtain their corresponding persistence diagrams. \hyperref[Figure 6.3]{\textcolor{blue}{Figure \ref{Figure 6.3}}} illustrates that the sample persistence diagram exhibits several short-lived $H_2$ features caused by contamination, whereas the trimmed persistence diagram isolates the dominant $H_2$ feature corresponding to the spherical void. In the sample persistence diagram, the dominant loop appears with a persistence interval of $(0.666, 0.962)$, corresponding to a persistence length of $0.296$. Such a short interval is difficult to separate from noise, so, by itself, it would not provide a reliable indicator of the true geometry. For each point, we compute the average pairwise distance to the rest of the sample points and asymmetric trimming is performed with trimming proportions $\alpha_1=0.2$ and $\alpha_2=0.04$. \hyperref[Figure 6.3]{\textcolor{blue}{Figure \ref{Figure 6.3}}} illustrates that the asymmetric trimming procedure (for $\alpha_1=0.2$ and $\alpha_2=0.04$) can recover the topological structure by suppressing spurious features introduced by contamination. 

\hyperref[Table 6.2]{\textcolor{blue}{Table \ref{Table 6.2}}} further summarizes how varying the trimming proportions alters the resulting topological summaries of $H_2$ feature by emphasizing the stabilizing role of trimming in recovering the core features of the data.
Even with marginal changes in trimming parameters, for instance, $\alpha_2$ rises from $0.01$ to $0.04$, the persistence length of the dominant $H_2$ feature extends from $0.327$ to $0.977$, nearly triples that of the untrimmed case. From a statistical point of view, this represents a substantial enhancement of the signal-to-noise ratio. The topological feature that was previously indistinguishable from noise becomes dominant and stable across scales. This demonstrates that removing even a very small fraction of the lowest average pairwise distance points can have a statistical impact. In contrast, increasing the upper trimming proportion $\alpha_1$, which targets the outermost points, produces only minor changes in persistence as long as it remains below $0.3$. When $\alpha_1$ becomes large (for example, $\alpha_1=0.4$), the highest persistence drops drastically and shifts to new birth–death coordinates, such as (0.891,0.994). This change suggests that excessive trimming disrupts the structural integrity of the sphere by replacing the genuine cavity with short-lived, artificial topological classes.

Overall, the results reveal a consistent pattern: moderate upper trimming ($\alpha_1 \approx 0.1 - 0.2$) combined with a small lower trimming proportion ($\alpha_2 \approx 0.04 - 0.05$) yields the most accurate recovery of the underlying spherical cavity. This demonstrates that with properly chosen trimming proportions, it is possible to recover the true topological features of the support of the underlying distribution by suppressing the contamination.
\begin{table}[h!]
\centering
\renewcommand{\thetable}{6.2}
\caption{Comparison of the highest persistence feature in $H_2$ of both the sample and the trimmed persistence diagrams of the spherical structure with clustered contamination}
\footnotesize
\begin{tabular}{ccccccc}
\toprule
$n$ & \multicolumn{2}{c}{Sample Persistence Diagram} & \multicolumn{2}{c}{Trimming Proportion} & \multicolumn{2}{c}{Trimmed Persistence Diagram} \\
\cmidrule(lr){2-3} \cmidrule(lr){4-5} \cmidrule(lr){6-7}
& Persistence Interval & Length & \quad\quad $\alpha_1$ & \quad$\alpha_2$ & Persistence Interval & Length \\
\midrule
400 & (0.6661787, 0.9615737)  & 0.295395 & \quad\quad 0.1 & \quad 0.01  & (0.6661787, 0.9936632) & 0.3274845 \\
     &  &  &  \quad\quad 0.2 & \quad 0.01  & (0.6661787, 0.9936632) & 0.3274845 \\
     &  &  & \quad\quad 0.3 & \quad 0.01  & (0.6661787, 0.9936632) & 0.3274845 \\
     &  &  & \quad\quad 0.4 & \quad 0.01  & (0.8908857, 0.9936632) & 0.1027776 \\ \\

     &  &  & \quad\quad 0.1 & \quad 0.02  & (0.6661787, 1.070676) & 0.4044972  \\
     &  &  & \quad\quad 0.2 & \quad 0.02  & (0.6661787, 1.070676) & 0.4044972 \\
     &  &  & \quad\quad 0.3 & \quad 0.02  & (0.6661787, 1.070676) & 0.4044972 \\
     &  &  & \quad\quad 0.4 & \quad 0.02  & (0.8975582, 1.070676) & 0.1731178 \\ \\
     
     &  &  & \quad\quad 0.1 & \quad 0.03  & (0.6661787, 1.174254) & 0.5080752 \\
     &  &  & \quad\quad 0.2 & \quad 0.03  & (0.6661787, 1.174254) & 0.5080752 \\
     &  &  & \quad\quad 0.3 & \quad 0.03  & (0.6661787, 1.174254) & 0.5080752 \\
     &  &  & \quad\quad 0.4 & \quad 0.03  & (0.9642748, 1.174254) & 0.209979 \\ \\

     &  &  & \quad\quad 0.1 & \quad 0.035  & (0.6661787, 1.223444) & 0.5572651 \\
     &  &  & \quad\quad 0.2 & \quad 0.035  & (0.6661787, 1.223444) & 0.5572651 \\
     &  &  & \quad\quad 0.3 & \quad 0.035  & (0.6661787, 1.223444) & 0.5572651 \\
     &  &  & \quad\quad 0.4 & \quad 0.035  & (1.160899, 1.223444) & 0.06254496 \\ \\

     &  &  & \quad\quad 0.1 & \quad 0.04  & (0.6661787, 1.64314) & 0.9769609 \\
     &  &  & \quad\quad 0.2 & \quad 0.04  & (0.6661787, 1.64314) & 0.9769609 \\
     &  &  & \quad\quad 0.3 & \quad 0.04  & (0.6661787, 1.64314) & 0.9769609 \\
     &  &  & \quad\quad 0.4 & \quad 0.04  & (1.592669, 1.652509)  & 0.05983999 \\ \\
     
     &  &  & \quad\quad 0.1 & \quad 0.05  & (0.6661787, 1.646691)  & 0.9805125 \\
     &  &  & \quad\quad 0.2 & \quad 0.05  & (0.6661787, 1.646691)  & 0.9805125 \\
     &  &  & \quad\quad 0.3 & \quad 0.05  & (0.6661787, 1.646691)  & 0.9805125 \\
     &  &  & \quad\quad 0.4 & \quad 0.05  & (1.592669, 1.652509)  & 0.05983999\\ \\

     &  &  & \quad\quad 0.1 & \quad 0.06  & (0.6661787, 1.646691)  & 0.9805125\\
     &  &  & \quad\quad 0.2 & \quad 0.06  & (0.6661787, 1.646691)  & 0.9805125\\
     &  &  & \quad\quad 0.3 & \quad 0.06  & (0.6661787, 1.646691)  & 0.9805125  \\
     &  &  & \quad\quad 0.4 & \quad 0.06  & (1.592669, 1.652509)  & 0.05983999 \\ 
\bottomrule
\end{tabular}
\end{table}\label{Table 6.2}
\section{Real-Data Analysis} \label{section 7}
In this section, we illustrate the practical utility and robustness of our proposed methodology through the analysis of a real-world dataset. The dataset concerns about the binding pocket in the three-dimensional structure of the streptavidin protein. The objective is to demonstrate how our theoretical developments provide meaningful insights into real-world scenarios, when data are contaminated and the underlying support exhibits non-trivial topological features.
\subsubsection*{Topological Characterisation of the Biotin-Binding Pocket in Streptavidin:}
In cellular biology, a protein's biological function is governed not only by its linear sequence of amino acids but also by its three-dimensional conformation. Such conformations often contain pockets (hollow regions) that act as binding sites for small ligands such as vitamins or drug molecules. Identifying these pockets is essential for understanding molecular function and for designing new medicines.

As a case study, we analyze the three-dimensional structure of streptavidin, a bacterial protein renowned for its extraordinarily high affinity to biotin (also known as Vitamin B7), see, e.g., \textcite{weber1989structural}, \textcite{ayan2022cooperative}. Due to this property, the streptavidin-biotin complex has become a cornerstone in biochemical research, biotechnology and targeted drug delivery. The atomic coordinates of streptavidin are publicly available in the Protein Data Bank (\url{https://www.rcsb.org/structure/7EK8}). Importantly, the structure was determined at ambient temperature in the absence of biotin, providing an ideal testbed for detecting structural pockets without ligand-induced distortions through topological methods. Although streptavidin is naturally tetrameric, we restrict attention to a single monomer for analysis from the tetramer to isolate the pocket geometry. The monomer retains the essential structural features of the binding pocket while reducing computational burden. To refine the monomer, we remove crystallographic waters and non-essential hetero-atoms, and restrict the representation to heavy atoms. The resulting monomeric structure contains 942 atoms, and each atom is represented by its Cartesian coordinates $(x,y,z) \in \mathbb{R}^3$ together with its chemical identity. These coordinates collectively form a three-dimensional point cloud that represents a streptavidin monomer. We have chosen this protein not only for its biological relevance but also because of the presence of distinct internal pockets that correspond to the biotin binding site. For clarity of analysis, we focus on a single monomer, which contains one functional binding site, thereby isolating the topological signature of the biotin-binding pocket without interference from inter-monomer interactions. Geometrically, the binding site is an internal cavity (a hollow region) inside the streptavidin into which biotin fits tightly. Topologically, such cavities appear as two-dimensional homology features ($H_2$), which can be detected and summarized using persistent homology.
However, structural data derived from X-ray crystallography or cryo-EM often introduces measurement noise, leading to false atomic coordinates in the point cloud. These contaminants act as outliers and can distort topological summaries. Since persistent homology is sensitive to outliers, such contaminants can induce unstable topological features in the analysis.

To address this issue, we apply our robust methodology based on an asymmetric trimming procedure parametrized by $(\alpha_1,\alpha_2)$. The upper trimming proportion $\alpha_1$ discards $\alpha_1\%$ of atoms with the largest average pairwise distances and, thereby, eliminates atoms far from the molecular core. The lower trimming proportion $\alpha_2$ removes atoms with the smallest average pairwise distances, which often correspond to tightly clustered or cavity-occupying atoms. Our proposed trimming procedure can isolate the core geometric structure of the protein while reducing the influence of both types of contamination.

After trimming, we construct a Vietoris–Rips filtration over the trimmed dataset, and compute the persistent homology up to dimension $k=2$. Our analysis emphasizes the $H_{2}$ homology class because the biotin-binding pocket is fundamentally a void. While $H_{0}$ and $H_{1}$ features correspond to clusters and loops, respectively, they do not represent solvent-accessible cavities. The resulting sample persistence diagram reveals many short-lived features with a persistent two-dimensional homology feature. This persistent $H_2$ feature corresponds to the biotin-binding cavity within the streptavidin protein. Its persistence across multiple scales reflects the structural stability of the binding pocket.

A comparison with the sample persistence diagram highlights the effect of contamination. The sample persistence diagram is dominated by numerous short-lived $H_2$ features due to extreme atomic coordinates, which obscure the biotin-binding pocket. This pocket appears with a persistence length of $\SI{1.274}{\angstrom}$ (persistence interval ($\SI{4.307}{\angstrom}, \SI{5.581}{\angstrom})$). Such a short interval is difficult to distinguish from noise, highlighting the need for robust inference.

\begin{figure}[h!]
        \centering
        \includegraphics[scale=0.25]{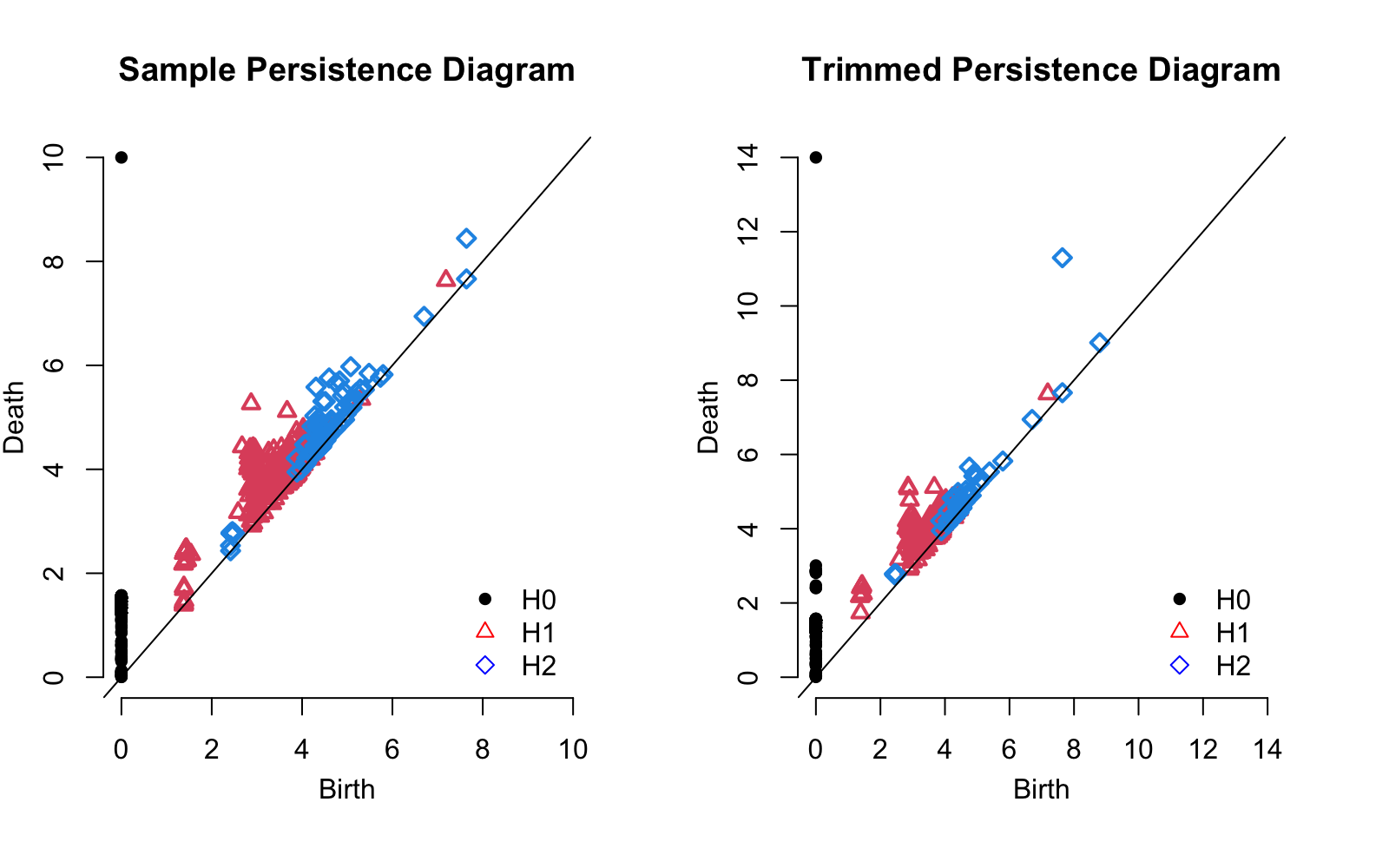}
        \caption{Comparison between Sample Persistence Diagram (left) and Trimmed Persistence Diagram (right) for $\alpha_1=0.3$ and $\alpha_2=0.05$.}
        \label{Figure 7.1}
\end{figure}

The asymmetric trimming results are summarized in  \hyperref[Table 7.1]{\textcolor{blue}{Table \ref{Table 7.1}}}. With $(\alpha_{1}, \alpha_{2})=(0.1,0.01)$, the persistence length of the cavity almost doubles $(\approx \SI{1.27}{\angstrom} \to ~\approx \SI{2.11}{\angstrom})$ relative to the untrimmed sample, indicating that trimming suppresses noisy points inside the cavity. Surprisingly at $(\alpha_{1}, \alpha_{2})=(0.1,0.02)$, persistence drops below the sample baseline, before rising steadily for $\alpha_{2}\geq 0.03$ ($\approx  \SI{1.56}{\angstrom}$ at $0.03$, $\approx \SI{2.45}{\angstrom}$ at $0.04$, $\approx \SI{3.66}{\angstrom}$ at $0.05$, and up to $\approx \SI{5.08}{\angstrom}$ at $0.08$). We interpret this non-monotone behaviour as a change in the identity of the dominant $H_2$ feature due to the boundary atoms essential to maintaining the cavity being removed. Beyond the empirical threshold $\alpha_{2}\approx 0.03$, trimming effectively removes cavity-occluding atoms, thereby restoring the dominant $H_2$ homology class and substantially increasing its persistence. In contrast, varying $\alpha_{1}$ between 0.10 and 0.40 by fixing $\alpha_{2}$ leaves identical persistence values across rows in \hyperref[Table 7.1]{\textcolor{blue}{Table \ref{Table 7.1}}}. This highlights that deleting a few atoms based on the largest average pairwise distances has a limited effect on the pocket’s $H_2$ persistence. \hyperref[Figure 7.1]{\textcolor{blue}{Figure \ref{Figure 7.1}}} shows that asymmetric trimming with $\alpha_1=0.3$ and $\alpha_2=0.05$ able to suppress contamination-induced spurious features and recovers the intrinsic $H_2$ cavity.

From geometric point of view, the expected biotin-binding pocket is approximately $\SI{5}{\angstrom}$ in diameter. Without trimming, the biotin-binding pocket is indistinguishable from short-lived $H_2$ features, whereas with trimming, it stands out distinctly as a stable, high-persistence $H_2$ feature. This demonstrates that our robust topological inference can be able to recover meaningful biological signals from noisy structural data.
\begin{table}[h!]
\centering
\renewcommand{\thetable}{7.1}
\caption{Summary of the highest persistence feature in the sample and trimmed persistence diagrams of sample size $n=942$.}
\footnotesize
\begin{tabular}{ccccccc}
\toprule
$n$ & \multicolumn{2}{c}{Sample Persistence Diagram} & \multicolumn{2}{c}{Trimming Proportion} & \multicolumn{2}{c}{Trimmed Persistence Diagram} \\
\cmidrule(lr){2-3} \cmidrule(lr){4-5} \cmidrule(lr){6-7}
& Persistence Interval & Length & \quad\quad $\alpha_1$ & \quad$\alpha_2$ & Persistence Interval & Length \\
\midrule
942 & (4.307264, 5.581416)  & 1.274152 & \quad\quad 0.1 & \quad 0.01  & (4.921503, 7.02886 ) & 2.107357 \\
     &  &  & \quad\quad 0.2 & \quad 0.01  & (4.921503, 7.02886 ) & 2.107357 \\
     &  &  & \quad\quad 0.3 & \quad 0.01  & (4.921503, 7.02886 ) & 2.107357 \\
     &  &  & \quad\quad 0.4 & \quad 0.01  & (4.921503, 7.02886 ) & 2.107357 \\ \\

942 & (4.307264, 5.581416)  & 1.274152 & \quad\quad 0.1 & \quad 0.02  & (4.598772, 5.757894) & 1.159122 \\
     &  &  & \quad\quad 0.2 & \quad 0.02  & (4.598772, 5.757894) & 1.159122 \\
     &  &  & \quad\quad 0.3 & \quad 0.02  & (4.598772, 5.757894) & 1.159122 \\ 
     &  &  & \quad\quad 0.4 & \quad 0.02  & (4.598772, 5.757894) & 1.159122 \\ \\

942 & (4.307264, 5.581416)  & 1.274152 & \quad\quad 0.1 & \quad 0.03  & (7.50245, 9.066539) & 1.564089 \\
     &  &  & \quad\quad 0.2 & \quad 0.03  & (7.50245, 9.066539) & 1.564089  \\
     &  &  & \quad\quad 0.3 & \quad 0.03  & (7.50245, 9.066539) & 1.564089  \\ 
     &  &  & \quad\quad 0.4 & \quad 0.03  & (7.50245, 9.066539) & 1.564089 \\ \\

942 & (4.307264, 5.581416)  & 1.274152 & \quad\quad 0.1 & \quad 0.04   & (7.6394, 10.09209) & 2.452687 \\
     &  &  & \quad\quad 0.2 & \quad 0.04  & (7.6394, 10.09209) & 2.452687 \\
     &  &  & \quad\quad 0.3 & \quad 0.04  & (7.6394, 10.09209) & 2.452687 \\
     &  &  & \quad\quad 0.4 & \quad 0.04  & (7.6394, 10.09209) & 2.452687 \\ \\

942 & (4.307264, 5.581416)  & 1.274152 & \quad\quad 0.1 & \quad 0.05   & (7.6394, 11.29894) & 3.659537 \\
     &  &  & \quad\quad 0.2 & \quad 0.05  & (7.6394, 11.29894) & 3.659537 \\
     &  &  & \quad\quad 0.3 & \quad 0.05  & (7.6394, 11.29894) & 3.659537 \\ 
     &  &  & \quad\quad 0.4 & \quad 0.05  & (7.6394, 11.29894) & 3.659537 \\ \\

942 & (4.307264, 5.581416)  & 1.274152 & \quad\quad 0.1 & \quad 0.06  & (7.6394, 11.73958) & 4.100176 \\
     &  &  & \quad\quad 0.2 & \quad 0.06  & (7.6394, 11.73958) & 4.100176 \\
     &  &  & \quad\quad 0.3 & \quad 0.06  & (7.6394, 11.73958) & 4.100176 \\ 
     &  &  & \quad\quad 0.4 & \quad 0.06  & (7.6394, 11.73958) & 4.100176 \\ \\

942 & (4.307264, 5.581416)  & 1.274152 & \quad\quad 0.1 & \quad 0.07  & (7.6394, 12.39076) & 4.751361 \\
     &  &  & \quad\quad 0.2 & \quad 0.07  & (7.6394, 12.39076) & 4.751361 \\
     &  &  & \quad\quad 0.3 & \quad 0.07  & (7.6394, 12.39076) & 4.751361 \\ 
     &  &  & \quad\quad 0.4 & \quad 0.07  & (7.6394, 12.39076) & 4.751361 \\ \\

942 & (4.307264, 5.581416)  & 1.274152 & \quad\quad 0.1 & \quad 0.08   & (7.6394, 12.71992) & 5.080516 \\
     &  &  & \quad\quad 0.2 & \quad 0.08   & (7.6394, 12.71992) & 5.080516 \\
     &  &  & \quad\quad 0.3 & \quad 0.08    & (7.6394, 12.71992) & 5.080516 \\
     &  &  & \quad\quad 0.4 & \quad 0.08   & (7.6394, 12.71992) & 5.080516 \\ 
\bottomrule
\end{tabular}
\label{Table 7.1}
\end{table}
\paragraph{Validation Beyond Persistence:} While persistence provides a robust statistical signal, we further validate our findings through a multi-evidence framework.
First, our analysis across a range of trimming parameters $(\alpha_{1}, \alpha_{2})$ (see \hyperref[Table 7.1]{\textcolor{blue}{Table \ref{Table 7.1}}}) demonstrates qualitative consistency of the dominant topological features, indicating that the recovered cavity is stable under small perturbations of the trimming parameters. This qualitative consistency supports the use of asymmetric trimming as a preliminary screening test for identifying meaningful topological cavities in noisy molecular point clouds. 
Second, biochemical knowledge derived from established structural and amino-acid sequence studies supports the interpretation that the identified void corresponds to the biotin-binding region in streptavidin. 
Although we do not perform new biochemical experiments, this concordance with prior biological knowledge provides an independent check that the cavity isolated by our statistical method corresponds to the region already known in streptavidin. 
Together, these lines of evidence guarantee the statistical stability and biological interpretability of our topological characterization of the streptavidin-biotin pocket, making it valuable as a preliminary screening stage prior to more detailed experimental investigations.
\section{Concluding Remarks}\label{section 8}
In this article, we introduced an asymmetric trimming framework for robust persistent homology based on average pairwise distances, which explicitly targets two types of contamination: outliers that lie outside the main data cloud, as well as the outliers lying inside that can obscure intrinsic topological features. The latter type of outliers is particularly pernicious as they can fill up genuine cavities, thereby shortening birth-death lifetimes in persistence diagrams and rendering true voids statistically indistinguishable from noise. By computing the average pairwise distance for each observation and trimming points from both the upper and lower tails based on the average pairwise distances, this procedure intuitively aims to isolate the core geometry of the underlying distribution by reducing the influence of noise and dense spurious clusters.
Persistent homology computed over this trimmed support yields diagrams that can reflect the essential topological structure. We further established uniform convergence in expected bottleneck distance of the trimmed persistence diagrams to their population analogues over the class of probability distributions under standard geometric and mass-regularity conditions, using a coupling argument to handle the dependence introduced by data-driven trimming. In addition, adaptive, data-driven rules for choosing the trimming proportions, together with finite-sample guarantees, improve the practical reliability of the method and make it easier to use in applications.

At the same time, this method faces several challenges: trimming choices can induce non-monotone transitions in persistence when features lie near boundaries, while excessive trimming can eliminate genuine structure. The exact computation of average pairwise distances also scales as $\mathcal{O}(n^2)$, which can be computationally expensive for large datasets. Moreover, under strong dependence, severe undersampling or in high-dimensional manifolds, the effect of trimming can be biased, and the coupling arguments and asymptotic guarantees may not hold. Hence, conclusions should be drawn with caution.

 The above issues suggest several directions to extend this work. On the methodological side, it would be natural to extend the framework to dependent data, irregular supports, and high-dimensional manifolds. On the computational side, scalable approximations of average pairwise distances, such as the $k$-nearest-neighbour technique, offer a promising direction to broaden applicability. Taken together, the asymmetric trimming methodology provides a simple, interpretable, and theoretically grounded approach to robust topological inference that can substantially improve the reliability of persistent homology in contaminated, real-world datasets.\\

 \noindent {\bf Acknowledgement:} The authors are grateful to Professor Wolfgang Polonik for insightful discussions and valuable suggestions that significantly influenced the theoretical development of this work. The authors also sincerely thank Sudipta Paul, a PhD student in the Department of Biological Sciences and Bioengineering, IIT Kanpur, for assistance and helpful input related to the data analysis.
 
\appendix
\section*{Appendix} \label{Appendix}
\subsection*{Proof of \hyperref[theorem 1]{\textcolor{blue}{Theorem \ref{theorem 1}}}}

Observe that in view of the stability theorem of persistence diagrams (see \cite{cohen2005stability}), we have
\small
\begin{equation*}
W_{\infty}\bigg(\widehat{\mathrm{\mathrm{Dgm}}}_{k}\Big(\mathscr{F}\big(\textbf{Supp}(\mathbb{P}_{n}^{\alpha_1,\alpha_2})\big)\Big), \mathrm{\mathrm{Dgm}}_{k}\Big(\mathscr{F}\big(\textbf{Supp}(\mathbb{P}^{\alpha_1,\alpha_2})\big)\Big)\bigg) \leq d_H\Big(\textbf{Supp}(\mathbb{P}_n^{\alpha_1,\alpha_2}),~ \textbf{Supp}(\mathbb{P}^{\alpha_1,\alpha_2})\Big),
\end{equation*}
\normalsize
where $d_H(\cdot,\cdot)$ denotes the Hausdorff distance. Hence, 

\small
\begin{align}\label{eq:4.1} 
    \mathbb{E}\Bigg[\bigg(\frac{(n-\lfloor{\alpha_1 n}\rfloor -\lfloor{\alpha_2 n}\rfloor)}{\log(n- \lfloor{\alpha_1 n}\rfloor - \lfloor{\alpha_2 n}\rfloor)}\bigg)^{\frac{1}{b}}W_{\infty} & \bigg( \widehat{ \mathrm{\mathrm{Dgm}}}_{k} \Big(\mathscr{F}\big(\textbf{Supp}(\mathbb{P}_{n}^{\alpha_1,\alpha_2}) \big)\Big), \mathrm{\mathrm{Dgm}}_{k}\Big(\mathscr{F}\big(\textbf{Supp}(\mathbb{P}^{\alpha_1,\alpha_2})\big)\Big)\bigg)\Bigg]\nonumber \\ 
    & \hspace{-2cm} \leq \mathbb{E}\Bigg[\underbrace{\bigg(\frac{(n-\lfloor{\alpha_1 n}\rfloor - \lfloor{\alpha_2 n}\rfloor)}{\log(n-\lfloor{\alpha_1 n}\rfloor - \lfloor{\alpha_2 n}\rfloor )}\bigg)^{\frac{1}{b}}d_H\Big(\textbf{Supp}(\mathbb{P}_n^{\alpha_1,\alpha_2}),~ \textbf{Supp}(\mathbb{P}^{\alpha_1,\alpha_2})\Big)}_{(\textbf{E1})}\Bigg].\tag{4.1}
\end{align}
\normalsize
Let us now work on (\textbf{E1})
uniformly over $\mathbb{P}\in \mathscr{P}_{a,b}(\mathcal{M})$.  Note that the compactness of $\mathcal{M}$  ensures that for $\varepsilon > 0$, there exists a finite $\varepsilon$-dense set $\mathbb{U}_{\mathbf{N}}=\{u_1,\cdots, u_{\mathbf{N}}\}\subseteq \textbf{Supp}(\mathbb{P}^{\alpha_1,\alpha_2})$, so the $\varepsilon$-covering number $\mathbf{N}:=\mathbf{N}\big(\varepsilon,\textbf{Supp}(\mathbb{P}^{\alpha_1,\alpha_2})\big)$ is finite. Therefore, the triangular inequality leads to
\small
\begin{align*}
    d_H\Big(\textbf{Supp}(\mathbb{P}_n^{\alpha_1,\alpha_2}),~ \textbf{Supp}(\mathbb{P}^{\alpha_1,\alpha_2})\Big) 
    &\leq d_H\Big(\textbf{Supp}(\mathbb{P}_n^{\alpha_1,\alpha_2}), \mathbb{U}_\mathbf{N}\Big) + d_H\Big(\mathbb{U}_\mathbf{N},~ \textbf{Supp}(\mathbb{P}^{\alpha_1,\alpha_2})\Big).
\end{align*}
\normalsize
 Further, since $\mathbb{U}_\mathbf{N}\subseteq \textbf{Supp}(\mathbb{P}^{\alpha_1,\alpha_2})$ is a minimal $\varepsilon$-dense set, it follows that $$d_H\Big(\mathbb{U}_\mathbf{N},~ \textbf{Supp}(\mathbb{P}^{\alpha_1,\alpha_2})\Big) \leq \varepsilon,$$ and note that $$d_H\big(\textbf{Supp}(\mathbb{P}_n^{\alpha_1,\alpha_2}), \mathbb{U}_\mathbf{N}\big) >\varepsilon \iff B_{\varepsilon}(u_i) ~\cap ~ \textbf{Supp}\left(\mathbb{P}_n^{\alpha_1,\alpha_2}\right)=\phi, ~ \text{for some}~~ u_i \in \mathbb{U}_\mathbf{N}.$$ Hence, we have 
 \small
\begin{align} \label{eq:4.2}
    \Pr \Big(d_H\big(\textbf{Supp}(\mathbb{P}_n^{\alpha_1,\alpha_2}),~ \textbf{Supp}(\mathbb{P}^{\alpha_1,\alpha_2})\big) > 2\varepsilon\Big)
    & \leq \overbrace{\Pr \Big(d_H\big(\textbf{Supp}(\mathbb{P}_n^{\alpha_1,\alpha_2}), \mathbb{U}_\mathbf{N}\big) > \varepsilon \Big)}^{(\textbf{E2})},\nonumber\\
    &= \underbrace{\Pr \Big[B_{\varepsilon}(u_i) ~\cap ~ \textbf{Supp}(\mathbb{P}_n^{\alpha_1,\alpha_2})=\phi, ~\text{for some}~ u_i \in \mathbb{U}_\mathbf{N} \Big]}_{(\textbf{E3})}. \tag{4.2}
\end{align}
\normalsize

Next, observe that due to data-dependent selection, the random variables in $\mathbb{X}_n^{\alpha_1,\alpha_2}$ are dependent random variables with non-identical distributions, although  $X_{i_j}$ has a marginal distribution $P_j \in \mathscr{P}_{a,b}(\mathcal{M)}$ for $ j=1,\cdots,(n-\lfloor{\alpha_1 n}\rfloor -\lfloor{\alpha_2 n}\rfloor)$. Therefore, using assumption (A3), there exists a collection of independent random variables $\mathbb{Z}_{n-\lfloor{\alpha_1 n}\rfloor - \lfloor{\alpha_2 n}\rfloor}$ = $\{Z_1,\cdots, Z_{n-\lfloor{\alpha_1 n}\rfloor - \lfloor{\alpha_2 n}\rfloor}\} $  defined on the same or on a richer probability space, and
for $j \in \{1,\cdots,n-\lfloor{\alpha_1 n}\rfloor - \lfloor{\alpha_2 n}\rfloor \}$, let us assume that  $\mu_j=\mathrm{Law}(X_{i_{j}})$ and $\nu_j=\mathrm{Law}(Z_{j})$.
Hence, the joint law of $(X_{i_j},Z_j)$ defines a coupling of the measures $\mu_j$ and $\nu_j$. Moreover, in view of the definition of coupling, we have
\small
\begin{align}\label{eq:4.3}
    \Pr\left(\mathbb{X}_n^{\alpha_1,\alpha_2} \neq \mathbb{Z}_{n-\lfloor{\alpha_1 n}\rfloor - \lfloor{\alpha_2 n}\rfloor}\right) &= \Pr \left(\exists ~\text{  $j \in \big\{1,\cdots,(n-\lfloor{\alpha_1 n}\rfloor -\lfloor{\alpha_2 n}\rfloor)\big\}$ such that}~ X_{i_j} \neq Z_j \right), \nonumber\\
    &\leq \sum_{j=1}^{n-\lfloor{\alpha_1 n}\rfloor -\lfloor{\alpha_2 n}\rfloor} \Pr \big(X_{i_j} \neq Z_j\big)~=~\delta_{n}. \quad (\text{by Boole's inequality)} \tag{4.3}
\end{align}  
\normalsize

Now, applying the union bound over all $u_i \in \mathbb{U}_N$ to (\textbf{E2}) in (\ref{eq:4.2}), we have 
\small
    \begin{align}\label{eq:4.4}
    \Pr \Big[d_H(\textbf{Supp}(\mathbb{P}_n^{\alpha_1,\alpha_2}), \mathbb{U}_\mathbf{N}) > \varepsilon \Big]&= \sum_{i=1}^\mathbf{N} \Pr\Big[B_{\varepsilon}(u_i) ~\cap ~ \textbf{Supp}(\mathbb{P}_n^{\alpha_1,\alpha_2})=\phi\Big], \nonumber \\
    &= \sum_{i=1}^\mathbf{N} \Pr\Big[B_{\varepsilon}(u_i) ~\cap ~ \textbf{Supp}(\mathbb{P}_n^{\alpha_1,\alpha_2})=\phi, ~\mathbb{X}_n^{\alpha_1,\alpha_2} = \mathbb{Z}_{n-\lfloor{\alpha_1 n}\rfloor-\lfloor{\alpha_2 n}\rfloor} \Big]~ +\nonumber \\
    & \quad \sum_{i=1}^\mathbf{N} \Pr\Big[B_{\varepsilon}(u_i) ~\cap ~ \textbf{Supp}(\mathbb{P}_n^{\alpha_1,\alpha_2})=\phi,  ~ \mathbb{X}_n^{\alpha_1,\alpha_2} \neq\mathbb{Z}_{n-\lfloor{\alpha_1 n}\rfloor -\lfloor{\alpha_2 n}\rfloor}\Big] \tag{4.4}\\
    &:=\textbf{E4}+\textbf{E5} \nonumber 
    \end{align} 
    \normalsize
    Next, we work on each term of the right-hand side of the equation (\ref{eq:4.4}):
       \begin{itemize}
            \item For \textbf{E4}, on the event $\mathbb{X}_n^{\alpha_1,\alpha_2}=\mathbb{Z}_{n-\lfloor{\alpha_1 n}\rfloor - \lfloor{\alpha_2 n}\rfloor}$, it follows that their supports coincide. Since $Z_1,\cdots, Z_{n-\lfloor{\alpha_1 n}\rfloor -\lfloor{\alpha_2 n}\rfloor}$ are constructed to be independent, we have
            \footnotesize
            \begin{align}
               \sum_{i=1}^\mathbf{N} \Pr\Big[B_{\varepsilon}(u_i) ~\cap ~ \textbf{Supp}(\mathbb{P}_n^{\alpha_1,\alpha_2})=\phi, ~\mathbb{X}_n^{\alpha_1,\alpha_2} = \mathbb{Z}_{n-\lfloor{\alpha_1 n}\rfloor -\lfloor{\alpha_2 n}\rfloor} \Big] &= \sum_{i=1}^\mathbf{N} \Pr\Big[B_{\varepsilon}(u_i) ~\cap ~ \textbf{Supp}(\mathbb{P}_n^Z)=\phi \Big],\nonumber\\
               &= \sum_{i=1}^\mathbf{N} \prod_{j=1}^{(n-\lfloor{\alpha_1 n}\rfloor -\lfloor{\alpha_2 n}\rfloor)} \Pr \big(Z_j \notin B_{\varepsilon}(u_i)\big) \label{eq:4.5} \tag{4.5}.
           \end{align}
           \normalsize
           \item For \textbf{E5}, due to the coupling discrepancy, we have 
           \footnotesize
           \begin{align}
               \sum_{i=1}^\mathbf{N} \Pr \Big[B_{\varepsilon}(u_i) ~\cap ~\textbf{Supp}(\mathbb{P}_n^{\alpha_1,\alpha_2})=\phi,  ~ \mathbb{X}_n^{\alpha_1,\alpha_2} \neq\mathbb{Z}_{n-\lfloor{\alpha_1 n}\rfloor -\lfloor{\alpha_2 n}\rfloor}\Big] &\leq \sum_{i=1}^\mathbf{N} \Pr \Big[\mathbb{X}_n^{\alpha_1,\alpha_2} \neq\mathbb{Z}_{n-\lfloor{\alpha_1 n}\rfloor -\lfloor{\alpha_2 n}\rfloor}\Big], \label{eq:4.6} \tag{4.6}\\
               &\leq \sum_{i=1}^\mathbf{N} \delta_n \nonumber.
           \end{align}
           \normalsize
       \end{itemize} 
Consequently, using (\ref{eq:4.4}), (\ref{eq:4.5}) and (\ref{eq:4.6}) lead to
\small
\begin{align}\label{eq:4.7}
    \Pr\Big[d_H(\textbf{Supp}(\mathbb{P}_n^{\alpha_1,\alpha_2}), \mathbb{U}_\mathbf{N}) > \varepsilon \Big]& \leq \sum_{i=1}^\mathbf{N} \prod_{j=1}^{(n-\lfloor{\alpha_1 n}\rfloor -\lfloor{\alpha_2 n}\rfloor)}\Pr\big(Z_j \notin B_{\varepsilon}(u_i)\big) + \sum_{i=1}^\mathbf{N} \delta_n,\nonumber
    \\& \leq  \sum_{i=1}^\mathbf{N} \prod_{j=1}^{(n-\lfloor{\alpha_1 n}\rfloor -\lfloor{\alpha_2 n}\rfloor)}\Big[1- \Pr \big(B_{\varepsilon}(u_i)\big)\Big] + \mathbf{N} ~\delta_n,\nonumber\\
        & \leq \mathbf{N}~\exp\Big\{-\big(n-\lfloor{\alpha_1 n}\rfloor -\lfloor{\alpha_2 n}\rfloor\big)\min\big\{1, a \varepsilon^b\big\}\Big\} + \mathbf{N} ~\delta_n.
        \tag{4.7}
    \end{align}
    \normalsize
    
Next, let $\mathbf{N}':=\mathbf{N}'(\varepsilon,\mathrm{A},d)$ be the $\varepsilon$-packing number of the support of the trimmed population measure $\mathbb{P}^{\alpha_1,\alpha_2}$, and hence, the minimal $\varepsilon$-covering number satisfies $\mathbf{N}(\varepsilon,\mathrm{A},d)\leq \mathbf{N}'(\varepsilon,\mathrm{A},d)$. Substituting it in (\ref{eq:4.7}), we obtain
\small
    \begin{align}\label{eq:4.8}
    \Pr \Big[d_H\big(\textbf{Supp}(\mathbb{P}_n^{\alpha_1,\alpha_2}), \mathbb{U}_\mathbf{N}\big) > \varepsilon \Big]& \leq \mathbf{N}'\Big[\exp\big\{-\big(n-\lfloor{\alpha_1 n}\rfloor -\lfloor{\alpha_2 n}\rfloor\big)\min\{1, a \varepsilon^b\}\big\} +  ~\delta_n\Big]. \tag{4.8}
    \end{align}
    \normalsize
Next, using (\ref{eq:2.2}) to $\textbf{Supp}(\mathbb{P}^{\alpha_1,\alpha_2})$ 
    and $\Pr \big(\textbf{Supp}(\mathbb{P}_n^{\alpha_1,\alpha_2})\big)\leq (1-\alpha_1-\alpha_2)$, we have
    \footnotesize
    \begin{align}\label{eq:4.9}
    \Pr \Big[d_H\big(\textbf{Supp}(\mathbb{P}_n^{\alpha_1,\alpha_2}), \mathbb{U}_\mathbf{N}\big) > \varepsilon \Big] & \leq \frac{2^b \Pr\big(\textbf{Supp}(\mathbb{P}^{\alpha_1,\alpha_2})\big)}{a \varepsilon^b} \Big[\exp\big\{-\big(n-\lfloor{\alpha_1 n}\rfloor -\lfloor{\alpha_2 n}\rfloor\big)\min\{1, a \varepsilon^b\}\big\} +  \delta_n\Big],\nonumber\\
    &\leq\frac{ 2^b (1-\alpha_1-\alpha_2)}{a \varepsilon^b} \Big[\exp\big\{-(n-\lfloor{\alpha_1 n}\rfloor -\lfloor{\alpha_2 n}\rfloor)\min\{1, a \varepsilon^b\}\big\} +  ~\delta_n\Big]. \tag{4.9}
    \end{align} 
    \normalsize
    Finally, in view of (\ref{eq:4.9}) and (\ref{eq:4.2}), we conclude:
    \footnotesize
    \begin{equation}\label{eq:4.10}
        \Pr \Big(d_H\big(\textbf{Supp}(\mathbb{P}_n^{\alpha_1,\alpha_2}),~ \textbf{Supp}(\mathbb{P}^{\alpha_1,\alpha_2})\big) > 2\varepsilon\Big) \leq \frac{ 2^b (1-\alpha_1-\alpha_2)}{a \varepsilon^b} \Big[\exp\big\{-(n-\lfloor{\alpha_1 n}\rfloor -\lfloor{\alpha_2 n}\rfloor)\min\{1, a \varepsilon^b\}\big\} +  ~\delta_n\Big]. \tag{4.10}
    \end{equation}
    \normalsize
    This completes the derivation of a finite-sample upper bound for the Hausdorff distance between the support of the trimmed empirical measure and that of the corresponding trimmed population measure.
    
For further simplification of the bound, the following two cases are considered. 

\vspace{0.25 in}
    
\noindent $\bullet$ {\large{\textbf{{\textbf{Case - 1:}} $\min\{1,a\varepsilon^b\} = a\varepsilon^b$}}}

\vspace{0.25 in}

\noindent This corresponds to the regime where the scale parameter $\varepsilon>0$ is sufficiently small such that  $a\varepsilon^b<1$.  Then inequality (\ref{eq:4.10}) yields,
\small
\begin{equation}\label{eq:4.11}
    \Pr \Big(d_H\big(\textbf{Supp}(\mathbb{P}_n^{\alpha_1,\alpha_2}),~ \textbf{Supp}(\mathbb{P}^{\alpha_1,\alpha_2})\big) > 2\varepsilon\Big) \leq \frac{2^b (1-\alpha_1-\alpha_2)}{a\varepsilon^b}\Big[ \exp\big\{-(n-\lfloor{\alpha_1 n}\rfloor -\lfloor{\alpha_2 n}\rfloor)~ a \varepsilon^b\big\}+\delta_n\Big]. \tag{4.11}
\end{equation}
\normalsize
Consider now $2 \varepsilon=s$  in (\ref{eq:4.11}), which leads to
\small
    \begin{align}\label{eq:4.12}
     \Pr \Bigg(\bigg(\frac{(n-\lfloor{\alpha_1 n}\rfloor -\lfloor{\alpha_2 n}\rfloor)}{\log(n-\lfloor{\alpha_1 n}\rfloor -\lfloor{\alpha_2 n}\rfloor)} \bigg)^\frac{1}{b}d_H &\big(\textbf{Supp}(\mathbb{P}_n^{\alpha_1,\alpha_2}), \textbf{Supp}(\mathbb{P}^{\alpha_1,\alpha_2})\big)\leq s\Bigg) \nonumber\\ & \hspace{-2.3cm}
        \geq 1-\frac{4^b (1-\alpha_1-\alpha_2)(n-\lfloor{\alpha_1 n}\rfloor -\lfloor{\alpha_2 n}\rfloor)}{a s^b \log(n-\lfloor{\alpha_1 n}\rfloor -\lfloor{\alpha_2 n}\rfloor)}\Big[ (n-\lfloor{\alpha_1 n}\rfloor -\lfloor{\alpha_2 n}\rfloor)^{-a 2^{-b}s^b}+\delta_n\Big]. \tag{4.12}
    \end{align}
    \normalsize
       Suppose that for any $s\in\mathbb{R}$, let us denote $$\hat{F_n} (s) = P\left[\bigg(\frac{(n-\lfloor{\alpha_1 n}\rfloor -\lfloor{\alpha_2 n}\rfloor)}{\log(n-\lfloor{\alpha_1 n}\rfloor -\lfloor{\alpha_2 n}\rfloor)} \bigg)^\frac{1}{b}~d_H\big(\textbf{Supp}(\mathbb{P}_n^{\alpha_1,\alpha_2}),~ \textbf{Supp}(\mathbb{P}^{\alpha_1,\alpha_2})\big)\leq s\right].$$ 
      Therefore, from (\ref{eq:4.12}), we obtain 
      \small
    \begin{equation}\label{eq:4.13}
        1- \hat{F_n}(s)\leq  \min\bigg( 1,~ \frac{4^b (1-\alpha_1-\alpha_2)(n-\lfloor{\alpha_1 n}\rfloor -\lfloor{\alpha_2 n}\rfloor)}{a s^b \log(n-\lfloor{\alpha_1 n}\rfloor -\lfloor{\alpha_2 n}\rfloor)}~ \big[ (n-\lfloor{\alpha_1 n}\rfloor -\lfloor{\alpha_2 n}\rfloor)^{-a 2^{-b}s^b}+\delta_n\big] \bigg), \tag{4.13}
    \end{equation}
    \normalsize
    for all $s>0$. Since $d_{H}(., .) \geq 0$, we have
    \small
\begin{align}\label{eq:4.14}
    \mathbb{E}\Bigg[\bigg(\frac{(n-\lfloor{\alpha_1 n}\rfloor -\lfloor{\alpha_2 n}\rfloor)}{\log(n-\lfloor{\alpha_1 n}\rfloor -\lfloor{\alpha_2 n}\rfloor)} \bigg)^\frac{1}{b}d_H\big(\textbf{Supp}(\mathbb{P}_n^{\alpha_1,\alpha_2}), \textbf{Supp}(\mathbb{P}^{\alpha_1,\alpha_2})\big)\Bigg] &=\int_{0}^{\infty} \Big[1-\hat{F_n}(s)\Big] ds, \nonumber
         \nonumber \\& \leq c+ \int_{c}^{\infty} \Big[1-\hat{F_n}(s)\Big] ds. \tag{4.14}
\end{align}
\normalsize
To ensure $\int_{c}^{\infty} \Big[1-\hat{F_n}(s)ds\Big]< \infty$, choose $c=c_{a,b}$ such that $1-a 2^{-b}s^b\leq 0 ,~ \text{for all}~ s\geq c_{a,b}\geq 0$, and hence, 
    \small
\begin{align}\label{eq:4.15}
        \mathbb{E}\Bigg[&\bigg(\frac{(n-\lfloor{\alpha_1 n}\rfloor -\lfloor{\alpha_2 n}\rfloor)}{\log(n-\lfloor{\alpha_1 n}\rfloor -\lfloor{\alpha_2 n}\rfloor)} \bigg)^\frac{1}{b}d_H\Big(\textbf{Supp}(\mathbb{P}_n^{\alpha_1,\alpha_2}),~ \textbf{Supp}(\mathbb{P}^{\alpha_1,\alpha_2})\Big)\Bigg]\nonumber\\& \quad\quad \quad\leq c_{a,b}+ \frac{4^b (1-\alpha_1-\alpha_2) (n-\lfloor{\alpha_1 n}\rfloor -\lfloor{\alpha_2 n}\rfloor)}{a ~ \log(n-\lfloor{\alpha_1 n}\rfloor -\lfloor{\alpha_2 n}\rfloor)} \int_{c_{a,b}}^{\infty} \frac{1}{s^b}\Big[ (n-\lfloor{\alpha_1 n}\rfloor -\lfloor{\alpha_2 n}\rfloor)^{-a 2^{-b}s^b}+\delta_n\Big] ds. \tag{4.15}
    \end{align}
    \normalsize
    Now notice that either $b > 1$ or $0 < b \leq 1$. These two cases are analysed separately in the following. 

    \vspace{0.25 in}
    
    $\blacksquare$~ \textbf{Sub-Case I \quad $b>1$:}
    Observe that in this case 
    $$ \int_{c_{a,b}}^{\infty} s^{-b}(n-\lfloor{\alpha_1 n}\rfloor -\lfloor{\alpha_2 n}\rfloor)^{-a 2^{-b}s^b} ds < \infty \quad \text{and}\quad \int_{c_{a,b}}^{\infty} s^{-b} ds < \infty,$$
for $c_{a,b}>0$. Hence, in view of $\delta_n \to 0$ as $n \to \infty$ and \eqref{eq:4.15}, we have  
 \small
    \begin{align}\label{eq:4.16}
       \mathbb{E}\Bigg[\bigg(\frac{(n-\lfloor{\alpha_1 n}\rfloor -\lfloor{\alpha_2 n}\rfloor)}{\log(n-\lfloor{\alpha_1 n}\rfloor -\lfloor{\alpha_2 n}\rfloor)} \bigg)^\frac{1}{b} & d_H\Big(\textbf{Supp}(\mathbb{P}_n^{\alpha_1,\alpha_2}), \textbf{Supp}(\mathbb{P}^{\alpha_1,\alpha_2})\Big)\Bigg] \nonumber\\&\leq c_{a,b}  + \frac{4^b(1-\alpha_1-\alpha_2)}{a~ (b-1)c_{a,b}^{b-1} } + \frac{4^b (1-\alpha_1-\alpha_2) (n-\lfloor{\alpha_1 n}\rfloor -\lfloor{\alpha_2 n}\rfloor) ~\delta_n}{a~ (b-1)c_{a,b}^{b-1}\log(n-\lfloor{\alpha_1 n}\rfloor -\lfloor{\alpha_2 n}\rfloor)}. \tag{4.16}
       \end{align}
\normalsize
Next, since $\delta_n =\mathcal{O}\Big(\frac{\log(n-\lfloor{\alpha_1 n}\rfloor -\lfloor{\alpha_2 n}\rfloor)}{(n-\lfloor{\alpha_1 n}\rfloor -\lfloor{\alpha_2 n}\rfloor)} \Big)$, as per assumption (A3), we have 
\small
    \begin{align}\label{eq:4.17}
         \sup_{\mathbb{P} ~\in ~\mathscr{P}_{a,b}(\mathcal{M)}} \mathbb{E}\Big[~d_H\big(\textbf{Supp}(\mathbb{P}_n^{\alpha_1,\alpha_2}),~ \textbf{Supp}(\mathbb{P}^{\alpha_1,\alpha_2})\big)\Big]&\leq c'_{a,b,\alpha_1,\alpha_2} \Bigg(\frac{\log(n-\lfloor{\alpha_1 n}\rfloor -\lfloor{\alpha_2 n}\rfloor)}{(n-\lfloor{\alpha_1 n}\rfloor -\lfloor{\alpha_2 n}\rfloor)} \Bigg)^\frac{1}{b}, \tag{4.17}
    \end{align}
    \normalsize
where the constant $c_{a,b,\alpha_1,\alpha_2}>0$.  This leads to the convergence rate stated in (\ref{eq:4.1}).

\vspace{0.25 in}

$\blacksquare$~ \textbf{Sub-Case II \quad $0<b \leq1$:}
    In this regime to investigate the behaviour of the bound, we focus on the integral term that appears in (\ref{eq:4.15}).
    \small
     \begin{equation}\label{eq:4.18}
         I_1= \int_{c_{a,b}}^{\infty}  \frac{1}{s^b}{\big[(n-\lfloor{\alpha_1 n}\rfloor -\lfloor{\alpha_2 n}\rfloor)^{1-a 2^{-b}s^b}+\delta_n\big]} ds. \tag{4.18}
     \end{equation}
     \normalsize
  To ensure that $I_1$ remains uniformly bounded for $0 < b \leq 1$, we have the following two facts.
  \begin{itemize}
    \item[\textbf{(Fact 1)}] For all $s>c_{a,b}$, impose $1-a 2^{-b}s^b \leq -s^{\frac{b}{2}}$. \\
    This inequality ensures that $(n-\lfloor{\alpha_1 n}\rfloor -\lfloor{\alpha_2 n}\rfloor)^{1-a 2^{-b}s^b} \to 0$ as $n \to \infty$ uniformly for $s \in [c_{a,b},\infty)$. The inequality is equivalent to
   $s^b(a 2^{-b} - s^{-\frac{b}{2}}) \geq 1$. As $s \to \infty$ we have $s^b\to \infty$ and $s^{-b} \to 0$ implying that the $s^b(a 2^{-b} - s^{-\frac{b}{2}}) \to \infty$. Therefore, the inequality becomes true for all large $s$.
    \item[\textbf{(Fact 2)}] The coupling discrepancy $\delta_n$ satisfies $\delta_n =o(1),$ as $n \to \infty$ (Assumption (A3)).
    \end{itemize}
In views of Facts 1 and 2 along with (\ref{eq:4.18}), we have
\small
    \begin{align}\label{eq:4.19}
        I_1&\leq \int_{c_{a,b}}^{\infty} \frac{\exp\big\{(1-a 2^{-b}s^b)\log (n-\lfloor{\alpha_1 n}\rfloor -\lfloor{\alpha_2 n}\rfloor)\big\}}{s^b} ~ds \quad \leq \int_{c_{a,b}}^{\infty} \exp\big\{-s^{\frac{b}{2}}\big\}\ ds \quad\leq \frac{2}{b} ~\Gamma\Big(\frac{2}{b}, c_{a,b}^\frac{b}{2}\Big), \tag{4.19}
    \end{align} 
\normalsize
where $\Gamma(s,x)=\int_{x}^{\infty} t^{s-1} ~e^{-t}dt$. Hence,  there exists a constant $c''_{a,b,\alpha_1,\alpha_2}>0$ depending on $a,b,\alpha_1$ and $\alpha_2$ such that
\small
    \begin{equation}\label{eq:4.20}
        \sup_{\mathbb{P}\in\mathscr{P}_{a,b}(\mathcal{M})} \mathbb{E}\Big[~d_H\big(\textbf{Supp}(\mathbb{P}_n^{\alpha_1,\alpha_2}),~ \textbf{Supp}(\mathbb{P}^{\alpha_1,\alpha_2})\big)\Big]\leq c''_{a,b,\alpha_1,\alpha_2}\bigg[\frac{\log(n-\lfloor{\alpha_1 n}\rfloor -\lfloor{\alpha_2 n}\rfloor)}{(n-\lfloor{\alpha_1 n}\rfloor -\lfloor{\alpha_2 n}\rfloor)} \bigg]^\frac{1}{b}, \tag{4.20}
    \end{equation} 
    \normalsize
in view of (\ref{eq:4.1}) using (\ref{eq:4.15}). 

\vspace{0.25 in}

$\bullet$~{\large{\textbf{Case - 2:} $\min\{1,a\varepsilon^b\} = 1$ and $b>1$}} 

    From the inequality (\ref{eq:4.10}), we obtain 
    \small
\begin{align}\label{eq:4.21}
    \Pr \Big(d_H(\textbf{Supp}(\mathbb{P}_n^{\alpha_1,\alpha_2}),~ \textbf{Supp}(\mathbb{P}^{\alpha_1,\alpha_2})) > 2\varepsilon\Big) \leq \frac{2^b (1-\alpha_1-\alpha_2)}{a\varepsilon^b}\big[ \exp\big\{-(n-\lfloor{\alpha_1 n}\rfloor -\lfloor{\alpha_2 n}\rfloor) \big\}+\delta_n\big]. \tag{4.21}
\end{align}
\normalsize
Next, arguing in the same way as in Case - 1, we obtain
    \small
    \begin{align}\label{eq:4.22}
        \mathbb{E}\Bigg[\bigg(\frac{(n-\lfloor{\alpha_1 n}\rfloor -\lfloor{\alpha_2 n}\rfloor)}{\log(n-\lfloor{\alpha_1 n}\rfloor -\lfloor{\alpha_2 n}\rfloor)} \bigg)^\frac{1}{b}~  d_H\Big(&\textbf{Supp}(\mathbb{P}_n^{\alpha_1,\alpha_2}),~ \textbf{Supp}(\mathbb{P}^{\alpha_1,\alpha_2})\Big)\Bigg] \nonumber\\
        & \leq c~+\frac{4^b (1-\alpha_1-\alpha_2)}{a(b-1)c^{b-1}}\cdot\frac{ (n-\lfloor{\alpha_1 n}\rfloor -\lfloor{\alpha_2 n}\rfloor)}{\log(n-\lfloor{\alpha_1 n}\rfloor -\lfloor{\alpha_2 n}\rfloor)}~\big(1+\delta_n\big).   \tag{4.22}
    \end{align}
    \normalsize
Finally, choosing $\delta_n =\mathcal{O}\Big(\frac{\log(n-\lfloor{\alpha_1 n}\rfloor -\lfloor{\alpha_2 n}\rfloor)}{(n-\lfloor{\alpha_1 n}\rfloor -\lfloor{\alpha_2 n}\rfloor)} \Big)$ (as in (A3)), one has a constant $c'''_{a,b,{\alpha_1,\alpha_2}}>0$ depending on $a,b,\alpha_1$ and $\alpha_2$ such that
\small
    \begin{equation}\label{eq:4.23}
        \mathbb{E}\bigg[~d_H\Big(\textbf{Supp}(\mathbb{P}_n^{\alpha_1,\alpha_2}),~ \textbf{Supp}(\mathbb{P}^{\alpha_1,\alpha_2})\Big)\bigg]\leq c'''_{a,b,{\alpha_1,\alpha_2}} \bigg(\frac{\log(n-\lfloor{\alpha_1 n}\rfloor -\lfloor{\alpha_2 n}\rfloor)}{(n-\lfloor{\alpha_1 n}\rfloor -\lfloor{\alpha_2 n}\rfloor)} \bigg)^\frac{1}{b}. \tag{4.23}
    \end{equation}
    \normalsize
Combining this with (\ref{eq:4.1}) yields the desired rate of convergence. 

Hence, the proof is complete.
\small
    \hfill$\Box$
    \normalsize 
\printbibliography
\addcontentsline{toc}{section}{References}
\end{document}